\documentclass[12pt]{article}

\usepackage[centertags]{amsmath}
\usepackage{amsfonts}
\usepackage{amssymb}
\usepackage{amsthm}
\usepackage{amsmath}
\usepackage{graphicx,color}
\usepackage{ulem}

\newcommand{\beq}{\begin{equation}}
\newcommand{\eeq}[1]{\label{#1}\end{equation}}
\newcommand{\bea}{\begin{eqnarray}}
\newcommand{\eea}[1]{\label{#1}\end{eqnarray}}
\newcommand{\sea}[2]{\begin{subequations}\label{#1}\begin{align} #2 \end{align}\end{subequations}}
\newcommand{\ba}[1]{\begin{align}\label{#1}}
\newcommand{\ea}{\end{align}}

\begin{document}
\setlength{\topmargin}{-1cm} \setlength{\oddsidemargin}{0cm}
\setlength{\evensidemargin}{0cm}
\begin{titlepage}
\begin{center}
{\Large \bf Higher-Spin Fermionic Gauge Fields and\\ Their Electromagnetic Coupling}

\vspace{20pt}

{\large Marc Henneaux, Gustavo Lucena G\'omez and Rakibur Rahman}

\vspace{12pt}

Physique Th\'eorique et Math\'ematique \& International Solvay Institutes\\
Universit\'e Libre de Bruxelles, Campus Plaine C.P. 231, B-1050 Bruxelles, Belgium

\end{center}
\vspace{20pt}

\begin{abstract}

We study the electromagnetic coupling of massless higher-spin fermions in flat space. Under the assumptions
of locality and Poincar\'e invariance, we employ the BRST-BV cohomological methods to construct consistent
parity-preserving off-shell cubic $1-s-s$ vertices. Consistency and non-triviality of the deformations not only
rule out minimal coupling, but also restrict the possible number of derivatives. Our findings are in complete
agreement with, but derived in a manner independent from, the light-cone-formulation results of Metsaev and the
string-theory-inspired results of Sagnotti-Taronna. We prove that any gauge-algebra-preserving vertex cannot
deform the gauge transformations. We also show that in a local theory, without additional dynamical higher-spin
gauge fields, the non-abelian vertices are eliminated by the lack of consistent second-order deformations.

\end{abstract}

\end{titlepage}

\newpage

\section{Introduction}

Consistent interacting theories of massless higher-spin fields in flat space are difficult to construct.
Severe restrictions arise from powerful no-go theorems~\cite{No-go,Porrati}, which prohibit, in Minkowski
space, minimal coupling to gravity, when the particle's spin $s\geq\tfrac{5}{2}$, as well as to
electromagnetism (EM), when $s\geq\tfrac{3}{2}$. However, these particles may still interact through
gravitational and EM  multipoles. Indeed, $\mathcal N=2$ SUGRA~\cite{SUGRA} allows massless gravitini to have
dipole and higher-multipole couplings, but forbids a non-zero $U(1)$ charge in flat space. Gravitational
and EM multipole interactions also show up, for example, as the $2-s-s$ and $1-s-s$ trilinear vertices constructed
in~\cite{2-3-3,2-s-s} for bosonic fields\,\footnote{In a local theory, some of these vertices may not be extended
beyond the cubic order~\cite{Porrati,2-3-3}.}.

These cubic vertices are but special cases of the general form $s-s'-s''$, that involves massless fields of
arbitrary spins. Metsaev's light-cone formulation~\cite{Metsaev_Boson,Metsaev} puts restrictions on the number
of derivatives in these vertices, and thereby provides a way of classifying them. For bosonic fields, while the
complete list of such vertices was given in~\cite{Cubic-general}, Noether procedure has been employed
in~\cite{Karapet} to explicitly construct off-shell vertices, which do obey the number-of-derivative restrictions.
Also, the tensionless limit of string theory gives rise to a set of cubic vertices, which are in one-to-one
correspondence with the ones of Metsaev, as has been noticed by Sagnotti-Taronna in~\cite{Taronna}, where
generating functions for off-shell trilinear vertices for both bosonic and fermionic fields were presented.

In this paper, we consider the coupling of a massless fermion of arbitrary spin to a $U(1)$ gauge field,
in flat spacetime of dimension $D\geq4$. Such a study is important in that fermionic fields are required by
supersymmetry, which plays a crucial role in string theory, which in turn involves higher-spin fields. This
fills a gap in the higher-spin literature, most of which is about bosons only (with~\cite{Metsaev,Taronna,FV}
among the exceptions). We do not consider mixed-symmetry fields, and restrict our attention to totally
symmetric Dirac fermions $\psi_{\mu_1...\mu_n}$, of spin $s=n+\tfrac{1}{2}$. For these fields, we
employ the powerful machinery of BRST-BV cohomological methods~\cite{BRST-BV} to construct systematically
consistent interaction vertices\,\footnote{The BRST-BV approach, in general, is very useful in obtaining
gauge-invariant manifestly Lorentz-invariant off-shell vertices for higher-spin fields~\cite{Tsulaia},
as has been emphasized recently in~\cite{Metsaev-BRST}.}, with the underlying assumptions of locality,
Poincar\'e invariance and conservation of parity, and without relying on other methods. The would-be
off-shell $1-s-s$ cubic vertices will complement their bosonic counterparts constructed in~\cite{2-s-s}.

The organization of the paper is as follows. We clarify our conventions and notations, and present our main
results in the next two Subsections. In Section~\ref{sec:BRST}, we briefly recall the BRST deformation
scheme~\cite{BRST-BV} for irreducible gauge theories. With this knowledge, we then move on to constructing
consistent off-shell $1-s-s$ vertices in the following three Sections. In particular, Section~\ref{sec:RS}
considers the massless Rarita-Schwinger field, while Section~\ref{sec:fivehalf} pertains to $s=\tfrac{5}{2}$,
and Section~\ref{sec:arbitrary} generalizes the results, rather straightforwardly, to arbitrary spin,
$s=n+\tfrac{1}{2}$. In Section~\ref{sec:noa1}, we prove an interesting property of the vertices under study:
an abelian $1-s-s$ vertex, i.e., a $1-s-s$  vertex that does not deform the original abelian gauge algebra, never
deforms the gauge transformations. Section~\ref{sec:compare} is a comparative
study of our results with those of Metsaev~\cite{Metsaev} and Sagnotti-Taronna~\cite{Taronna}, where we
explicitly show their equivalence. Section~\ref{sec:2nd} investigates whether there are obstructions to the
existence of second-order deformations corresponding the non-abelian vertices, i.e., if they are consistent
beyond the cubic order. We conclude with some remarks in Section~\ref{sec:remarks}. Two appendices are added
to present some useful technical details, much required for the bulk of the paper.

\subsection{Conventions \& Notations}

We work in Minkowski spacetime with mostly positive metric. The Clifford algebra is
$\{\gamma^\mu,\gamma^\nu\}=+2\eta^{\mu\nu}$, and $\gamma^{\mu\,\dagger}=\eta^{\mu\mu}\gamma^\mu$.
The Dirac adjoint is defined as $\bar{\psi}_\mu=\psi^\dagger_\mu\gamma^0$. The $D$-dimensional
Levi-Civita tensor, $\epsilon_{\mu_1\mu_2...\mu_D}$, is normalized as $\epsilon_{01...(D-1)}=+1$.
We define $\gamma^{\mu_1....\mu_n}=\gamma^{[\mu_1}\gamma^{\mu_2}...\gamma^{\mu_n]}$, where
the notation $[i_1...i_n]$ means totally antisymmetric expression in all the indices $i_1,...,i_n$
with the normalization factor $\tfrac{1}{n!}$. The totally symmetric expression $(i_1...i_n)$
has the same normalization. We use the slash notation: $\gamma^\mu Q_\mu\equiv\displaystyle{\not{\!Q}}$.

The curvature for the spin-1 field is its 1-curl, $F_{\mu\nu}=\partial_\mu A_\nu-\partial_\nu A_\mu$, which
is just the EM field strength. Its contraction with two $\gamma$-matrices, $\gamma^{\mu\nu}F_{\mu\nu}$, is
denoted as ${\displaystyle\not{\!F\,}}$. Similarly, the curvature for the spin-$\tfrac{3}{2}$ field is given
by the 1-curl, $\Psi_{\mu\nu}=\partial_\mu\psi_\nu-\partial_\nu\psi_\mu$. For arbitrary spin $s=n+\tfrac{1}{2}$,
we have a totally symmetric rank-$n$ tensor-spinor $\psi_{\mu_1...\mu_n}$, whose curvature is a rank-$2n$
tensor-spinor, $\Psi_{\mu_1\nu_1|\mu_2\nu_2|...|\mu_n\nu_n}$, defined as the $n$-curl,
\begin{equation*} \Psi_{\mu_1\nu_1|\mu_2\nu_2|...|\mu_n\nu_n}\equiv\left[...\left[\,\left[\partial_{\mu_1}...
\partial_{\mu_n}\psi_{\nu_1...\nu_n}-(\mu_1\leftrightarrow\nu_1)\right]-(\mu_2\leftrightarrow\nu_2)\right]...
\right]-(\mu_n\leftrightarrow\nu_n).\end{equation*} This is the Weinberg curvature tensor~\cite{Curvature},
and we discuss more about it in Appendix~\ref{sec:Curvature}.

More generally, the rank-$n$ field $\psi_{\nu_1...\nu_n}$ can have an $m$-curl, for any $0\leq m\leq n$,
\begin{equation*} \psi^{(m)}_{\mu_1\nu_1|...|\mu_m\nu_m\|\nu_{m+1}...\nu_n}\equiv\left[...\left[\,\left[\partial
_{\mu_1}...\partial_{\mu_m}\psi_{\nu_1...\nu_n}-(\mu_1\leftrightarrow\nu_1)\right]-(\mu_2\leftrightarrow\nu_2)
\right]...\right]-(\mu_m\leftrightarrow\nu_m).\end{equation*} When $m=n$, this is nothing but the
curvature tensor, $\psi^{(n)}_{\mu_1\nu_1|...|\mu_n\nu_n}=\Psi_{\mu_1\nu_1|...|\mu_n\nu_n}$, whereas
$m=0$ corresponds to the original field itself, $\psi^{(0)}_{\nu_1...\nu_n}=\psi_{\nu_1\nu_2...\nu_n}$.

The Fronsdal tensor for the fermionic field~\cite{Fronsdal} will be denoted as $\mathcal S_{\mu_1...\mu_n}$, so that
\beq \mathcal S_{\mu_1...\mu_n}=i\left[\displaystyle{\not{\!\partial\,}}\psi_{\mu_1...\mu_n}
-n\partial_{(\mu_1}\displaystyle{\not{\!\psi\,}}_{\mu_2...\mu_n)}\right].\eeq{Fronsdal}

The symbol ``$\approx$'' will mean off-shell equivalence of two vertices, whereas ``$\sim$'' will stand for
equivalence in the transverse-traceless gauge (up to an overall factor).

\subsection{Results}

\begin{itemize}
 \item For massless fermions, we present a cohomological proof of the well-known fact that minimal EM coupling in flat
 space is ruled out for $s\geq\frac{3}{2}$~\cite{Porrati,Metsaev}.
 \item We find restrictions on the possible number of derivatives in a cubic $1-s-s$ vertex, with $s=n+\tfrac{1}{2}$.
 There are only three allowed values: $2n-1, 2n$, and $2n+1$. This is in complete agreement with the results of
 Metsaev~\cite{Metsaev}.
 \item The ($2n-1$)-derivative vertex is non-abelian $-$ the only one that deforms the gauge symmetry.
 With $F^{+\mu\nu}\equiv F^{\mu\nu}+\tfrac{1}{2}\gamma^{\mu\nu\alpha\beta}F_{\alpha\beta}$,
 we find that the off-shell vertex is simply
 \begin{equation*} s=\tfrac{3}{2}:~ \bar{\psi}_\mu F^{+\mu\nu}\psi_\nu,~~...\,,~~s=n+\tfrac{1}{2}:~
 \bar{\psi}^{(n-1)}_{\mu_1\nu_1|...|\mu_{n-1}\nu_{n-1}\|\,\mu}F^{+\mu}_{~~~\nu}\psi^{(n-1)\mu_1\nu_1|...|\mu_{n-1}
 \nu_{n-1}\|\,\nu}.\end{equation*} We see that the $(n-1)$-curl of the fermionic field appears in the vertex.
 \item The $2n$-derivative vertex is abelian. It exists only for $D>4$, and is gauge invariant up to a total
 derivative. It involves the curvature tensor, and takes the form
 \begin{equation*} s=n+\tfrac{1}{2}:~(\bar{\Psi}_{\mu_1\nu_1|\mu_2\nu_2|...|\mu_n\nu_n}\gamma^{\mu_1\nu_1\alpha_1
 \beta_1\lambda}\Psi_{\alpha_1\beta_1|}^{~~~~~~\mu_2\nu_2|...|\mu_n\nu_n})A_\lambda\,.\end{equation*}
 It can be interpreted as a Chern-Simons term. To see this, let us start with the spin-$\tfrac{3}{2}$ case.
 For any choice of spinor indices $a$, $b$, the expression $\bar{\Psi}_a\wedge\,\Psi^b$ defines a closed $4$-form,
 $d\left(\bar{\Psi}_a\wedge\,\Psi^b\right)=0$. Here $\Psi^b=\frac{1}{2}\Psi^b_{\alpha\beta}\,dx^\alpha\wedge
 dx^\beta$ and a similar expression holds for $\bar{\Psi}_a$. The associated Chern-Simons $5$-form  $$\bar{\Psi}_a
 \wedge\,\Psi^b\wedge A=\tfrac{1}{4}\bar{\Psi}_{a\vert\mu_1\nu_1}\Psi_{\alpha_1\beta_1}^{b}A_\lambda\,dx^{\mu_1}
 \wedge dx^{\nu_1} \wedge dx^{\alpha_1}\wedge dx^{\beta_1}\wedge dx^{\lambda},$$ with $A=A_\lambda dx^\lambda$, is
 therefore gauge invariant up to the exterior derivative of a $4$-form.  Replacing $dx^{\mu_1} \wedge dx^{\nu_1}
 \wedge dx^{\alpha_1} \wedge dx^{\beta_1} \wedge dx^{\lambda}$ by $(\gamma^{\mu_1\nu_1\alpha_1\beta_1\lambda})^
 a_{~\,b}$ and summing over the spinor indices give a scalar, which is gauge invariant up to a total divergence.
 This understanding of the $2n$-derivative vertex explains why it exists only in $D\geq 5$. For higher-spin fields,
 the curvatures are not exterior forms since they are described by mixed-symmetry Young tableaux. However, the
 contracted expression $$ \bar{\Psi}_a\wedge_\lrcorner\,\Psi^b=\tfrac{1}{4}\bar{\Psi}_{a \vert \mu_1\nu_1|\mu_2\nu_2|
 ...|\mu_n\nu_n}\Psi_{\alpha_1\beta_1|}^{b~~~~~\mu_2\nu_2|...|\mu_n\nu_n} dx^{\mu_1} \wedge dx^{\nu_1} \wedge
 dx^{\alpha_1} \wedge dx^{\beta_1}$$ is a closed $4$-form, and the construction proceeds then in the same way.
 \item The ($2n+1$)-derivative vertex, which is the highest-derivative one, is a 3-curvature term (Born-Infeld type).
 \item The non-abelian cubic vertices generically get obstructed, in a local theory, at second order deformation.
 In special cases, such vertices may extend beyond the cubic order, if additional dynamical higher-spin gauge fields
 are present.
\end{itemize}

\section{The BRST Deformation Scheme}\label{sec:BRST}

As pointed out in~\cite{BRST-BV}, one can reformulate the classical problem of introducing consistent
interactions in a gauge theory in terms of the BRST differential and the BRST cohomology. The advantage is
that the search for all possible consistent interactions becomes systematic, thanks to the cohomological approach.
Obstructions to deforming a gauge-invariant action also become related to precise cohomological classes of the
BRST differential. In what follows, we briefly explain the BRST deformation scheme.

Let us consider an irreducible gauge theory of a collection of fields $\{\phi^i\}$, with $m$ gauge invariances,
$\delta_{\varepsilon}\phi^i=R^i_\alpha\varepsilon^\alpha, \alpha=1,2,...,m$. Corresponding to each gauge parameter
$\varepsilon^\alpha$, one introduces a ghost field $\mathcal C^\alpha$, with the same algebraic symmetries but
opposite Grassmann parity ($\epsilon$). The original fields and ghosts are collectively called fields, denoted
by $\Phi^A$. The configuration space is further enlarged by introducing, for each field and ghost, an antifield
$\Phi^*_A$, that has the same algebraic symmetries (in its indices when $A$ is a multi-index) but opposite
Grassmann parity.

In the algebra generated by the fields and antifields, we introduce
two gradings: the pure ghost number ($pgh$) and the antighost number ($agh$). The former is non-zero only for the
ghost fields. In particular, for irreducible gauge theories, $pgh(\mathcal C^\alpha)=1$, while $pgh(\phi^i)=0$ for
any original field. The antighost number, on the other hand, is non-zero only for the antifields $\Phi^*_A$.
Explicitly, $agh(\Phi^*_A)=pgh(\Phi^A)+1,~agh(\Phi^A)=0=pgh(\Phi^*_A)$. The ghost number ($gh$) is another grading,
defined as $gh=pgh-agh$.

On the space of fields and antifields, one defines an odd symplectic structure \beq (X,Y)\equiv\frac{\delta^RX}
{\delta\Phi^A}\frac{\delta^LY}{\delta\Phi^*_A}-\frac{\delta^RX}{\delta\Phi^*_A}\frac{\delta^LY}{\delta\Phi^A},
\eeq{antibracket} called the antibracket\,\footnote{This definition gives $\left(\Phi^A,\Phi^*_B\right)=\delta^A_B$,
which is \emph{real}.  Because a field and its antifield have opposite Grassmann parity, it follows that if $\Phi^A$
is real, $\Phi^*_B$ must be purely imaginary, and vice versa.}. It satisfies the graded Jacobi identity.

The original gauge-invariant
action $S^{(0)}[\phi^i]$ is then extended to a new action $S[\Phi^A,\Phi^*_A]$, called the master action, that
includes terms involving ghosts and antifields, \beq S[\Phi^A,\Phi^*_A]=S^{(0)}[\phi^i]+\phi^*_iR^i_\alpha
\mathcal C^\alpha+~...~,\eeq{S_0} which, by virtue of the Noether identities and the higher-order gauge structure equations,
satisfies the classical master equation \beq (S,S)=0.\eeq{master} In other words, the master action $S$ incorporates
compactly all the consistency conditions pertaining to the gauge transformations. This also plays role as the
generator of the BRST differential $\mathfrak s$, which is defined as
\beq \mathfrak{s}X\equiv(S,X).\eeq{brst1}

Notice that $S$ is BRST-closed, as a simple consequence of the master equation. From the properties of the
antibracket, it also follows that $\mathfrak s$ is nilpotent, \beq \mathfrak s^2=0.\eeq{brst2} Therefore, the
master action $S$ belongs to the cohomology of $\mathfrak s$ in the space of local functionals of the fields,
antifields, and their finite number of derivatives.

As we know, the existence of the master action $S$ as a solution of the master equation is completely equivalent
to the gauge invariance of the original action $S^{(0)}[\phi^i]$. Therefore, one can reformulate the problem of
introducing consistent interactions in a gauge theory as that of deforming the solution $S$ of the master equation.
Let $S$ be the solution of the \emph{deformed} master equation, $(S,S)=0$. This must be a deformation of the
solution $S_0$ of the master equation of the \emph{free} gauge theory, in the deformation parameter $g$,
\beq S=S_0+gS_1+g^2S_2+\mathcal O(g^3).\eeq{brst5} The master equation for $S$ splits, up to $\mathcal O(g^2)$,
into \bea &(S_0,S_0)=0,&\label{brst6.1}\\&(S_0,S_1)=0,&\label{brst6.2}\\&(S_1,S_1)=-2(S_0,S_2).&\eea{brst6.3}
Eqn.~(\ref{brst6.1}) is fulfilled by assumption, and in fact $S_0$ is the generator of the BRST differential
for the \emph{free} theory, which we will denote as $s$. Thus, Eqn.~(\ref{brst6.2}) translates to
\beq s S_1=0,\eeq{brst1st} i.e., $S_1$ is BRST-closed. If the first-order \emph{local} deformations are given by
$S_1=\int a$, where $a$ is a top-form of ghost number 0, then one has the cocycle condition \beq sa+db=0.
\eeq{cocycle0} Non-trivial deformations therefore belong to $H^0(s|d)$ $-$ the cohomology of the zeroth-order
BRST differential $s$, modulo total derivative $d$, at ghost number 0. Now, if one makes an antighost-number
expansion of the local form $a$, it stops at $agh=2$~\cite{2-s-s,BBH,gravitons}, \beq a=a_0+a_1+a_2, \qquad
agh(a_i)=i=pgh(a_i).\eeq{brst7} For \emph{cubic} deformations $S_1=\int a$, it is indeed impossible to construct
an object with $agh>2$~\cite{2-s-s}. The result is however more general and holds in fact also for higher order
deformations, as it follows from the results of~\cite{BBH,gravitons,N=1SUGRA}.

The significance of the various terms is worth recalling. $a_0$ is the deformation of the Lagrangian, while $a_1$
and $a_2$ encode information about the deformations of the gauge transformations and the gauge algebra
respectively~\cite{BRST-BV}. Thus, if $a_2$ is not trivial, the algebra of the gauge transformations is deformed and
becomes non-abelian. On the other hand, if $a_2 = 0$ (up to redefinitions), the algebra remains abelian to first order
in the deformation parameter. In that case, if $a_1$ is not trivial, the gauge transformations are deformed (remaining
abelian), while if $a_1 = 0$ (up to redefinitions), the gauge transformations remain the same as in the undeformed case.

The various gradings are of relevance as $s$ decomposes into the sum of the Koszul-Tate differential,
$\Delta$, and the longitudinal derivative along the gauge orbits, $\Gamma$: \beq s=\Delta+\Gamma. \eeq{brst3}
$\Delta$ implements the equations of motion (EoM) by acting only on the antifields. It decreases the antighost
number by one unit while keeping unchanged the pure ghost number. $\Gamma$ acts only on the original fields and
produces the gauge transformations. It increases the pure ghost number by one unit without modifying the antighost
number. Accordingly, all three $\Delta$, $\Gamma$ and $s$ increase the ghost number by one unit,
$gh(\Delta)=gh(\Gamma)=gh(s)=1$. Note that $\Delta$ and $\Gamma$ are nilpotent and anticommuting,
\beq \Gamma^2=\Delta^2=0,\qquad\Gamma\Delta+\Delta\Gamma=0.\eeq{brst4}

Given the expansion~(\ref{brst7}) and the decomposition~(\ref{brst3}), the cocycle condition~(\ref{cocycle0}) yields
the following cascade of relations, that a consistent deformation must obey \bea &\Gamma a_2=0,\label{cocycle1}&\\&
\Delta a_2+\Gamma a_1+db_1=0,&\label{cocycle2}\\&\Delta a_1+\Gamma a_0+db_0=0,&\eea{cocycle3} where
$agh(b_i)=i,~pgh(b_i)=i+1$. Note that $a_2$ has been chosen to be $\Gamma\text{-closed}$, instead of
$\Gamma\text{-closed}$ modulo $d$, as is always possible~\cite{BBH}.

We now analyze the conditions under which $a_2$ and $a_1$ are non-trivial.
\begin{itemize}
\item Non-triviality of the deformation of the gauge algebra: The highest-order term $a_2$ will be trivial
(i.e., removable by redefinitions) if and only if one can get rid of it by adding to $a$ an $s$-exact term modulo $d$,
$sm + dn$. Expanding $m$ and $n$ according to the antighost number, and taking into account the fact that $m$ and $n$
also stop at $agh=2$ since they are both cubic, one finds that $a_2$ is trivial if and only if $a_2=\Gamma
m_2 + dn_2$. We see that the cohomology of $\Gamma$ modulo $d$ plays an important role. The cubic vertex will deform
the gauge algebra if and only if $a_2$ is a non-trivial element of the cohomology of $\Gamma$ modulo $d$. Otherwise,
can always choose $a_2=0$, and $a_1=\Gamma\text{-closed}$~\cite{BBH}. Note that since $a_2$ is a cocycle of the cohomology
of $\Gamma$ modulo $d$, which can be chosen to be $\Gamma$-closed~\cite{BBH}, one can investigate the general
form of $a_2$ by studying the elements in the cohomology of $\Gamma$ that are not $d$-exact.
\item Non-triviality of the deformation of the gauge transformations: We now assume $a_2 = 0$. In this case,  $a_1$ can
be chosen to be a non-trivial cocycle of $\Gamma$. The vertex deforms the gauge transformations unless $a_1$ is
$\Delta$-exact modulo $d$, $a_1 = \Delta m_2 + dn_1$, where $m_2$ can be assumed to be
invariant~\cite{BBH,gravitons,N=1SUGRA}. In that instance, one can remove $a_1$, and so one can take $a_0$ to be
$\Gamma$-closed modulo $d$: the vertex only deforms the action without deforming the gauge transformations.
The cohomology of $\Delta$ is also relevant in that the Lagrangian deformation $a_0$ is $\Delta$-closed, whereas trivial
interactions are given by $\Delta$-exact terms.
\end{itemize}

Finally, while the graded Jacobi identity for the antibracket renders $(S_1,S_1)$ BRST-closed, the second-order
consistency condition~(\ref{brst6.3}) requires that this actually be $s$-exact:
\beq (S_1,S_1)=-2s S_2.\eeq{brst2nd} This condition determines whether or not, in a local theory, a consistent
first-order deformation gets obstructed at the second order. Such obstructions are controlled by the local
BRST cohomology group $H^1(s|d)$.

\subsection{The Cohomology of $\Gamma$}\label{sec:Gamma}

In this Subsection we present some facts about the cohomology of $\Gamma$, which will be very useful in the latter
parts of the paper. Details will be relegated to Appendix~\ref{sec:Gamma1}.

The system of gauge fields under consideration consists of a photon $A_\mu$ and a rank-$n$ spinor-tensor $\psi_{\mu_1...\mu_n}$.
Corresponding to them, there will be a Grassmann-odd ghost field $C$ and a Grassmann-even rank-$(n-1)$ spinorial ghost
$\xi_{\mu_1...\mu_{n-1}}$, which obeys $\displaystyle{\not{\!\xi}}_{\mu_1...\mu_{n-2}}=0$. The set of antifields is
$\Phi^*_A=\left\{A^{*\mu}, C^*, \bar{\psi}^{*\mu_1...\mu_n}, \bar{\xi}^{*\mu_1...\mu_{n-1}}\right\}$. We note that
the cohomology of $\Gamma$ is isomorphic to the space of local functions depending on

\begin{itemize}
 \item The curvatures $\left\{F_{\mu\nu}, \Psi_{\mu_1\nu_1|\mu_2\nu_2|...|\mu_n\nu_n}\right\}$, and their derivatives.
 \item The antifields  $\left\{A^{*\mu}, C^*, \bar{\psi}^{*\mu_1...\mu_n}, \bar{\xi}^{*\mu_1...\mu_{n-1}}\right\}$,
 and their derivatives.
 \item The undifferentiated ghosts $\left\{C, \xi_{\mu_1...\mu_{n-1}}\right\}$, and the $\gamma$-traceless part of all
 possible curls of the spinorial ghost $\left\{\xi^{(m)}_{\mu_1\nu_1|...|\mu_m\nu_m\|\nu_{m+1}...\nu_{n-1}},\, m\leq n-1
 \right\}$.
 \item The Fronsdal tensor $\mathcal S_{\mu_1...\mu_n}$, and its derivatives.
\end{itemize}

The derivative of the curl, $\partial_{\nu_n}\xi^{(m)}_{\mu_1\nu_1|...|\mu_m\nu_m\|\,\nu_{m+1}...\nu_{n-1}}$,
is of special interest. If and only if symmetrized w.r.t. the indices $\{\nu_{m+1}, ..., \nu_n\}$, does this
quantity become $\Gamma$-exact: \begin{equation*} \partial_{(\nu_n}\xi^{(m)\mu_1\nu_1|...|\mu_m\nu_m\|}
_{~~~~~~~~~~~~~~~~~~\,\nu_{m+1}...\nu_{n-1})}=\tfrac{1}{n-m}\,\Gamma\,\psi^{(m)\mu_1\nu_1|...|\mu_m\nu_m\|}
_{~~~~~~~~~~~~~~~~~~\,\nu_{m+1}...\nu_n},\qquad 0\leq m\leq n-1.\end{equation*} Incidentally, for the $(n-1)$-curl
one has $\partial_{\nu_n}\xi^{(n-1)}_{\mu_1\nu_1|...|\mu_{n-1}\nu_{n-1}}=\Gamma\,\psi^{(n-1)}_{\mu_1\nu_1|...|
\mu_{n-1}\nu_{n-1}\|\,\nu_n}~$.

\section{EM Coupling of Massless Spin 3/2}\label{sec:RS}

In this Section we construct parity-preserving off-shell $1-\tfrac{3}{2}-\tfrac{3}{2}$ vertices by employing the BRST-BV
cohomological methods. The spin-$\tfrac{3}{2}$ system is simple enough so that one can implement the BRST deformation
scheme with ease, while it captures many non-trivial features that could serve as guidelines as one moves on to higher spins.

The starting point is the free theory, which contains a photon $A_\mu$ and a massless Rarita-Schwinger field $\psi_\mu$,
described by the action \beq S^{(0)}[A_\mu,\psi_\mu]=\int d^Dx\left[-\tfrac{1}{4}F_{\mu\nu}^2-i\bar{\psi}_\mu\gamma^{\mu\nu\rho}
\partial_\nu\psi_\rho\right],\eeq{rs1} which enjoys two abelian gauge invariances: \beq \delta_\lambda A_\mu=\partial_\mu
\lambda,\qquad\delta_\varepsilon\psi_\mu=\partial_\mu\varepsilon.\eeq{rs2} For the Grassmann-even bosonic gauge parameter
$\lambda$, we introduce the Grassmann-odd bosonic ghost $C$. Corresponding to the Grassmann-odd fermionic gauge parameter
$\varepsilon$, we have the Grassmann-even fermionic ghost $\xi$. Therefore, the set of fields becomes
\beq \Phi^A=\{A_\mu, C, \psi_\mu, \xi\}.\eeq{rs3} For each of these fields, we introduce an antifield with the same algebraic
symmetries in its indices but opposite Grassmann parity. The set of antifields is \beq \Phi^*_{A}=\{A^{*\mu}, C^*, \bar{\psi}^{*\mu},
\bar{\xi}^*\}.\eeq{rs4} Now we construct the \emph{free} master action $S_0$, which is an extension of the original
gauge-invariant action~(\ref{rs1}) by terms involving ghosts and antifields. Explicitly, \beq S_0=\int d^Dx\left[-\tfrac{1}{4}
F_{\mu\nu}^2-i\bar{\psi}_\mu\gamma^{\mu\nu\rho}\partial _\nu\psi_\rho+A^{*\mu}\partial_\mu C - (\bar{\psi}^{*\mu}
\partial_\mu\xi-\partial_\mu\bar{\xi}\psi^{*\mu})\right].\eeq{rs5} Notice that the antifields appear as sources
for the ``gauge'' variations, with gauge parameters replaced by corresponding ghosts. It is easy to verify
that~(\ref{rs5}) indeed solves the master equation $(S_0,S_0)=0$. The different gradings and Grassmann parity
of the various fields and antifields, along with the action of $\Gamma$ and $\Delta$ on them, are given in Table 1.

\begin{table}[ht]
\caption{Properties of the Various Fields \& Antifields ($n=1$)}
\vspace{6pt}
\centering
\begin{tabular}{c c c c c c c}
\hline\hline
$Z$ &$\Gamma(Z)$~~~&~~~$\Delta(Z)$~~~&$pgh(Z)$ &$agh(Z)$ &$gh(Z)$ &$\epsilon(Z)$\\ [0.5ex]
\hline
$A_\mu$ & $\partial_\mu C$ & 0 & 0 & 0 & 0 & 0\\
$C$ & 0 & 0 & 1 & 0 & 1 & 1\\
$A^{*\mu}$ & 0 & $-\partial_\nu F^{\mu\nu}$ & 0 & 1 & $-1$ & 1\\
$C^*$ & 0 & $-\partial_\mu A^{*\mu}$ & 0 & 2 & $-2$ & 0\\ \hline
$\psi_\mu$ & $\partial_\mu\xi$ & 0 & 0 & 0 & 0 & 1\\
$\xi$ & 0 & 0 & 1 & 0 & 1 & 0\\
$\bar{\psi}^{*\mu}$ & 0 & $-\tfrac{i}{2}\bar{\Psi}_{\alpha\beta}\gamma^{\alpha\beta\mu}$ & 0 & 1 & $-1$ & 0\\
$\bar{\xi}^*$ & 0 & $\partial_\mu\bar{\psi}^{*\mu}$ & 0 & 2 & $-2$ & 1\\
\hline\hline
\end{tabular}
\end{table}
\vspace{6pt}

For the spin-$\tfrac{3}{2}$ field the Fronsdal tensor is $\mathcal S_\mu=i\left[\displaystyle{\not{\!\partial\,}}
\psi_{\mu}-\partial_{\mu}\displaystyle{\not{\!\psi\,}}\right]=-i\gamma^\nu\Psi_{\mu\nu}$, i.e., the $\gamma$-trace of
the curvature. The cohomology of $\Gamma$ is isomorphic to the space of functions of
\begin{itemize}
 \item The undifferentiated ghosts $\{C, \xi\}$,
 \item The antifields $\{A^{*\mu}, C^*, \bar{\psi}^{*\mu},\bar{\xi}^*\}$ and their derivatives,
 \item The curvatures $\{F_{\mu\nu}, \Psi_{\mu\nu}\}$ and their derivatives.
\end{itemize}

\subsection{Gauge-Algebra Deformation}

The next step is to consider, for the first-order deformation, the most general form of $a_2$ $-$ the term with $agh=2$,
that contains information about the deformation of the gauge algebra. $a_2$ must satisfy $\Gamma a_2=0$, and be Grassmann even
with $gh(a_2)=0$. Besides, we require that $a_2$ be a parity-even Lorentz scalar. Then, the most general possibility
is \beq a_2=-g_0C\left(\bar{\xi}^*\xi+\bar{\xi}\xi^*\right)-g_1C^*\bar{\xi}\xi,\eeq{rs6} which is a linear combination of
two independent terms: one contains \emph{both} the bosonic ghost $C$ and the fermionic ghost $\xi$, while the other contains
only $\xi$ but \emph{not} $C$. The former one potentially gives rise to minimal coupling, while the latter could produce
dipole interaction. This can be understood by first noting that the corresponding Lagrangian deformation, $a_0$, is obtained
through the consistency cascade~(\ref{cocycle1})--(\ref{cocycle3}). From the action of $\Gamma$ and $\Delta$ on the fields
and antifields, it is then easy to see that the respective $a_0$ would contain no derivative and one derivative respectively.

\subsection{Deformation of Gauge Transformations}

Next, we would like to see if $a_2$ can be lifted to certain $a_1$, i.e., with the given $a_2$, if one could solve
Eqn.~(\ref{cocycle2}) to find an $a_1$. Indeed, one finds that\,\footnote{Here one also needs
the relations $\Delta\xi^*=-\partial_\mu\psi^{*\mu},\,\Gamma\bar{\psi}_\mu=-\partial_\mu\bar{\xi}$, which follow
from Table 1.} \bea \Delta a_2&=&+g_0C\left[(\partial_\mu\bar{\psi}^{*\mu})\xi-\bar{\xi}(\partial_\mu\psi^{*\mu})\right]
+g_1(\partial_\mu A^{*\mu})\bar{\xi}\xi\nonumber\\&=&-g_0\left[\bar{\psi}^{*\mu}\partial_\mu(C\xi)-\partial_\mu(C\bar{\xi})
\psi^{*\mu}\right]-g_1A^{*\mu}\partial_\mu(\bar{\xi}\xi)+d(...)\nonumber\\&=&-\Gamma\left[g_0(\bar{\psi}^{*\mu}\psi_\mu
+\bar{\psi}_\mu\psi^{*\mu})C+g_0(\bar{\psi}^{*\mu}A_\mu\xi-\bar{\xi}A_\mu\psi^{*\mu})+g_1A^{*\mu}(\bar{\psi}_\mu\xi-\bar{\xi}
\psi_\mu)\right]+d(...).\nonumber\eea{x} Therefore, in view of Eqn.~(\ref{cocycle2}), one must have
\beq a_1=g_0\left[\bar{\psi}^{*\mu}(\psi_\mu C+\xi A_\mu)+\text{h.c.}\right]+g_1A^{*\mu}(\bar{\psi}_\mu\xi-\bar{\xi}\psi_\mu)
+\tilde{a}_1,\qquad \Gamma\tilde{a}_1=0,\eeq{rs8} where the ambiguity, $\tilde{a}_1$, belongs to the cohomology of $\Gamma$.
Its most general form will be \beq \tilde{a}_1=\left[\bar{\psi}^{*\mu}X_{\mu\nu\rho}\Psi^{\nu\rho}\right]C
+\left[\bar{\psi}^{*\mu}Y_{\mu\nu\rho}F^{\nu\rho}+\bar{\Psi}^{\mu\nu}Z_{\mu\nu\rho}A^{*\rho}\right]\xi+\text{h.c.},
\eeq{ambiguity} where $X, Y$ and $Z$ may contain derivatives and spinor indices.

\subsection{Lagrangian Deformation}

We note that $\Delta a_1$ must be $\Gamma$-closed modulo $d$, since
\beq \Gamma(\Delta a_1)=\Delta(-\Gamma a_1)=\Delta\left[\Delta a_2+d(...)\right]=d(...).\eeq{rs8.5}
Condition~(\ref{cocycle3}), however, requires that $\Delta a_1$ be $\Gamma$-exact modulo $d$. The $\Delta$-variation
of neither of the unambiguous pieces in $a_1$ is $\Gamma$-exact modulo $d$, and the non-trivial part must be killed
by $\Delta\tilde{a}_1$, if~(\ref{cocycle3}) holds at all. But such a cancelation is impossible for the first piece,
i.e., the would-be minimal coupling, simply because $\tilde{a}_1$ contains too many derivatives. Thus, minimal coupling
is ruled out, and we must set $g_0=0$. Then, we have
\beq \Delta a_1=-\Gamma(g_1\bar{\psi}_\mu F^{\mu\nu}\psi_\nu)-\tfrac{1}{2}g_1F^{\mu\nu}(\bar{\Psi}_{\mu\nu}\xi-\bar{\xi}
\Psi_{\mu\nu})+\Delta\tilde{a}_1+d(...).\eeq{rs9}
The second term on the right hand side is in the cohomology of $\Gamma$, and must be canceled by $\Delta\tilde{a}_1$.
To see if this is possible or not, we make use of the identity
\beq \eta^{\mu\nu|\alpha\beta}\equiv\tfrac{1}{2}\left(\eta^{\mu\alpha}\eta^{\nu\beta}-\eta^{\mu\beta}\eta^{\nu\alpha}\right)
=\tfrac{1}{2}\gamma^{\mu\nu}\gamma^{\alpha\beta}-2\gamma^{[\mu}\eta^{\nu][\alpha}\gamma^{\beta]}-\tfrac{1}{2}
\gamma^{\mu\nu\alpha\beta},\eeq{id0} to rewrite the term as
\bea F^{\mu\nu}(\bar{\Psi}_{\mu\nu}\xi-\bar{\xi}\Psi_{\mu\nu})&=&+\tfrac{1}{2}\left(\displaystyle{\not{\!\bar{\Psi}}}
\displaystyle{\not{\!F}}-4\bar{\Psi}_{\mu\nu}\gamma^{\mu}F^{\nu\rho}\gamma_\rho\right)\xi-\tfrac{1}{2}\bar{\xi}
\left(\displaystyle{\not{\!F}}\displaystyle{\not{\!\Psi}}-4\gamma_{\mu}F^{\mu\alpha}\gamma^\beta\Psi_{\alpha\beta}\right)
\nonumber\\&&-\tfrac{1}{2}\left(\bar{\Psi}_{\mu\nu}\gamma^{\mu\nu\alpha\beta}F_{\alpha\beta}\xi-\bar{\xi}F_{\mu\nu}
\gamma^{\mu\nu\alpha\beta}\Psi_{\alpha\beta}\right)\nonumber\\\nonumber\\&=&+\tfrac{1}{2}\left(\displaystyle{\not{\!\bar
{\Psi}}}\displaystyle{\not{\!F}}-4\bar{\Psi}_{\mu\nu}\gamma^{\mu}F^{\nu\rho}\gamma_\rho\right)\xi-\tfrac{1}{2}\bar{\xi}
\left(\displaystyle{\not{\!F}}\displaystyle{\not{\!\Psi}}-4\gamma_{\mu}F^{\mu\alpha}\gamma^\beta\Psi_{\alpha\beta}\right)
\nonumber\\&&+\Gamma\left(\bar{\psi}_\mu\gamma^{\mu\nu\alpha\beta}F_{\alpha\beta}\psi_\nu\right)+d(...).\eea{rs10}
Notice that, we have rendered the second line in the first step $\Gamma$-exact modulo $d$, by virtue of the Bianchi
identity, $\partial_{[\mu}F_{\nu\rho]}=0$. We plug Eqn.~(\ref{rs10}) into~(\ref{rs9}) to obtain
\bea \Delta a_1&=&-\Gamma(g_1\bar{\psi}_\mu F^{+\mu\nu}\psi_\nu)+\Delta\tilde{a}_1+d(...)\nonumber\\
&&-\tfrac{1}{4}g_1\left[\left(\displaystyle{\not{\!\bar{\Psi}}}\displaystyle{\not{\!F}}-4\bar{\Psi}_{\mu\nu}
\gamma^{\mu}F^{\nu\rho}\gamma_\rho\right)\xi-\bar{\xi}\left(\displaystyle{\not{\!F}}\displaystyle{\not{\!\Psi}}
-4\gamma_{\mu}F^{\mu\alpha}\gamma^\beta\Psi_{\alpha\beta}\right)\right].\eea{rs11} Now, the most important point
is that, the terms in the second line of the above expression are $\Delta$-exact, such that it is consistent to
set \beq \Delta\tilde{a}_1=\tfrac{1}{4}g_1\left[\left(\displaystyle{\not{\!\bar{\Psi}}}\displaystyle{\not{\!F}}
-4\bar{\Psi}_{\mu\nu}\gamma^{\mu}F^{\nu\rho}\gamma_\rho\right)\xi-\bar{\xi}\left(\displaystyle{\not{\!F}}
\displaystyle{\not{\!\Psi}}-4\gamma_{\mu}F^{\mu\alpha}\gamma^\beta\Psi_{\alpha\beta}\right)\right].\eeq{rs12}
This is tantamount to setting
\beq \tilde{a_1}=ig_1\left[\bar{\psi}^{*\mu}\gamma^\nu F_{\mu\nu}-\tfrac{1}{2(D-2)}\displaystyle{\not{\!\bar{\psi}^*}}
\displaystyle{\not{\!F}}\right]\xi+\text{h.c.},\eeq{rs13} which, of course, is in the cohomology of
$\Gamma$. Then, Eqn.~(\ref{rs11}) reduces to \beq \Delta a_1=-\Gamma(g_1\bar{\psi}_\mu F^{+\mu\nu}\psi_\nu)+d(...),
\eeq{rs14} so that we have a consistent Lagrangian deformation $a_0$. To summarize, we have
\beq a_0=g_1\bar{\psi}_\mu F^{+\mu\nu}\psi_\nu,\qquad a_1=g_1A^{*\mu}(\bar{\psi}_\mu\xi-\bar{\xi}\psi_\mu)+\tilde{a}_1,
\qquad a_2=-g_1C^*\bar{\xi}\xi.\eeq{rs15}

\subsection{Abelian Vertices}\label{sec:a1a0}

Now that we have exhausted all the possibilities for $a_2$, any other vertex can only have a trivial $a_2$.
In this case, as we will show in Section~\ref{sec:noa1}, one can always choose to write a vertex as the photon
field $A_\mu$ contracted with a gauge-invariant current $j^\mu$, \beq a_0=j^\mu A_\mu,\qquad \Gamma j^\mu=0,
\eeq{rs17} where the divergence of the current is $\Delta$-exact: \beq \partial_\mu j^\mu=\Delta M,\qquad
\Gamma M=0,\eeq{rs18} so that one has $a_1=MC$. If, however, $M$ happens to be $\Delta$-exact modulo $d$
in the space of invariants, one can add a $\Delta$-exact term in $a_0$, so that the new current is identically
conserved~\cite{BBH}. In the latter case, the vertex does not deform the gauge symmetry at all.

Now the most general vertex of the form~(\ref{rs17}) contains the current \beq j^\lambda=\bar{\Psi}_{\mu\nu}\,
X^{\mu\nu\alpha\beta\lambda}\,\Psi_{\alpha\beta},\eeq{rs19} whose divergence is required to obey the
condition~(\ref{rs18}). Here $X$ may contain Dirac matrices as well as derivatives. It is not difficult to see if
$X$ contains more than one derivatives, $a_0$ is $\Delta$-exact modulo $d$, i.e., trivial. First, if $X$ contains
the Laplacian, $\Box$, the contribution is always $\Delta$-exact, by the EoM $\Box\Psi_{\mu\nu}=0$.
We can also forgo the Dirac operator, $\displaystyle{\not{\!\partial}}$, because by using the relation
$\displaystyle{\not{\!\partial}}\gamma^\mu=2\partial^\mu-\gamma^\mu\displaystyle{\not{\!\partial}}$, one can always make
$\displaystyle{\not{\!\partial}}$ act on the curvature to get $\Delta$-exact terms, thanks to the EoM
$\displaystyle{\not{\!\partial}}\Psi_{\mu\nu}=0$. Therefore, any derivative contained in
$X^{\mu\nu\alpha\beta\lambda}$ must carry one of the five indices. Given the EoM $\partial^\mu\Psi_{\mu\nu}=0$,
the antisymmetry of the field strength $\Psi_{\mu\nu}$, and the commutativity of ordinary derivatives,
the only potentially non-trivial way to have more-than-one derivatives is \beq a_0=\left(\bar{\Psi}_{\mu\alpha}
\overset\leftarrow{\partial}_{\nu}\,\gamma^\lambda\,\partial^\mu\Psi^{\alpha\nu}\right)A_\lambda.\eeq{rs20}
But algebraic manipulations show that this vertex is actually $\Delta$-exact modulo $d$, i.e., trivial. To see
this, we use $\Psi^{\alpha\nu}=\partial^\alpha\psi^\nu-\partial^\nu\psi^\alpha$, and rewrite~(\ref{rs20}) as
\begin{equation*} a_0=\left[\bar{\Psi}_{\mu\alpha}\overset\leftarrow{\partial}_{\nu}\,\gamma^\lambda\,\partial^\mu
\partial^\alpha\psi^\nu-\tfrac{1}{2}\bar{\Psi}_{\mu\alpha}\overset\leftarrow{\partial}_{\nu}\,\gamma^\lambda\,
\partial^\nu\Psi^{\mu\alpha}\right]A_\lambda.\end{equation*} While the first term is identically zero, in the second
term, one can use the 3-box rule, $2\partial_\mu X\partial^\mu Y=\Box(XY)-X(\Box Y)-(\Box X)Y$, so that
\begin{equation*} a_0=-\tfrac{1}{4}\left[\Box\left(\bar{\Psi}_{\mu\alpha}\,\gamma^\lambda\,\Psi^{\mu\alpha}\right)
-\left(\Box\bar{\Psi}_{\mu\alpha}\right)\,\gamma^\lambda\,\Psi^{\mu\alpha}-\bar{\Psi}_{\mu\alpha}\,\gamma^\lambda\,
\left(\Box\Psi^{\mu\alpha}\right)\right]A_\lambda.\end{equation*} Here, the last two terms are $\Delta$-exact,
whereas in the first term a double integration by parts gives $\Box A_\lambda$, which is equal to $\partial_\lambda
(\partial\cdot A)$ by the photon EoM. Then, one is left with \begin{equation*} a_0=-\tfrac{1}{4}\left(\bar{\Psi}_
{\mu\alpha}\,\gamma^\lambda\,\Psi^{\mu\alpha}\right)\partial_\lambda(\partial\cdot A)+\Delta\text{-exact}+d(...).
\end{equation*} Upon integrating by parts w.r.t. $\partial_\lambda$, this indeed becomes $\Delta$-exact modulo $d$,
\beq a_0=\left(\bar{\Psi}_{\mu\alpha}\overset\leftarrow{\partial}_{\nu}\,\gamma^\lambda\,\partial^\mu
\Psi^{\alpha\nu}\right)A_\lambda=\Delta\text{-exact}+d(...).\eeq{rs21}

The only possibilities are therefore that $X$ contains either no derivative or one derivative. For the former case, we
have the candidate $X^{\mu\nu\alpha\beta\lambda}=-2\eta^{\mu\nu|\alpha\beta}\gamma^\lambda$. This gives
\beq M=-4i\bar{\Psi}_{\mu\nu}\partial^\mu\left(\psi^{*\nu}-\tfrac{1}{D-2}\gamma^\nu\displaystyle\not{\!\psi}\right)
-\text{h.c.},\eeq{M2der} which is obviously gauge invariant: $\Gamma M=0$. However, explicit computation easily
shows that $M$ is actually $\Delta$-exact modulo $d$. Therefore, one can render the current identically conserved
by adding a $\Delta$-exact term. In fact, in view of identity~(\ref{id0}), our candidate $j^\mu$ is
\beq j^\mu=\tfrac{1}{2}\bar{\Psi}_{\mu\nu}\left(\gamma^{\mu\nu\alpha\beta}\gamma^\lambda+\gamma^\lambda
\gamma^{\mu\nu\alpha\beta}\right)\Psi_{\alpha\beta}+\Delta\text{-exact}.\eeq{j2der}
Then, it is clear from the identity \beq \tfrac{1}{2}\gamma^{\mu\nu\alpha\beta}\gamma^\lambda+\tfrac{1}{2}
\gamma^\lambda\gamma^{\mu\nu\alpha\beta}=\gamma^{\mu\nu\alpha\beta\lambda},\eeq{id3} that our 2-derivative
vertex is actually off-shell equivalent ($\approx$) to \beq a_0\approx\left(\bar{\Psi}_{\mu\nu}\,
\gamma^{\mu\nu\alpha\beta\lambda}\,\Psi_{\alpha\beta}\right)A_\lambda.\eeq{rs22} This vertex does not deform
the gauge symmetry, and is gauge invariant up to a total derivative. Note that the vertex does not exist
in $D=4$, because of the presence of $\gamma^{\mu\nu\alpha\beta\lambda}$. This is in complete agreement with
Metsaev's results~\cite{Metsaev}.

Finally, we are left with the possibility of having just one derivative in $X$, which would correspond to a
3-derivative vertex. The only candidate is $X^{\mu\nu\alpha\beta\lambda}=\tfrac{1}{2}\eta^{\mu\nu|\alpha\beta}
\overset\leftrightarrow\partial\,^\lambda$, which is equivalent to $-\tfrac{1}{4}\gamma^{\mu\nu\alpha\beta\lambda}
\overset\leftrightarrow\partial\,^\lambda$, up to $\Delta$-exact terms, thanks to the identity~(\ref{id0}). We have
\beq a_0=\tfrac{1}{2}\left(\bar{\Psi}_{\mu\nu}\,\eta^{\mu\nu|\alpha\beta}\,\overset\leftrightarrow\partial\,^\lambda\,
\Psi_{\alpha\beta}\right)A_\lambda=\tfrac{1}{2}\left(\bar{\Psi}_{\mu\nu}\partial^\lambda\Psi^{\mu\nu}-\bar{\Psi}_{\mu\nu}
\overset\leftarrow\partial\,^\lambda\,\Psi^{\mu\nu}\right)A_\lambda.\eeq{rs23} In this case too, our candidate current
reduces on-shell to an identically conserved one, so that the vertex actually does not deform the gauge symmetry.
To see this, we use the Bianchi identity $\partial^\lambda\Psi^{\mu\nu}=-\partial^\mu\Psi^{\nu\lambda}+\partial^\nu
\Psi^{\mu\lambda}$, to write the vertex as \begin{equation*}a_0=\left(-\bar{\Psi}_{\mu\nu}\partial^\mu\Psi^{\nu\lambda}
+\bar{\Psi}_\nu^{~\lambda}\overset\leftarrow\partial_\mu\Psi^{\mu\nu}\right)A_\lambda.\end{equation*} Thanks to the EoM
$\partial^\mu\Psi_{\mu\nu}=0$, up to $\Delta$-exact terms, the current reduces to the total derivative of a fermion bilinear,
which is identically conserved: \beq a_0\approx2\partial_\nu\left(\bar{\Psi}_\alpha^{~[\mu}\Psi^{\nu]\alpha}\right)A_\mu.
\eeq{rs24} Upon integration by parts, this is just a 3-curvature term (Born-Infeld type),
\beq a_0\approx\bar{\Psi}_{\mu\alpha}\Psi^\alpha_{~\nu}F^{\mu\nu}.\eeq{rs25}
This exhausts all possible $1-\tfrac{3}{2}-\tfrac{3}{2}$ vertices. Below we present a summary table.
\begin{table}[ht]
\caption{Summary of $1-\tfrac{3}{2}-\tfrac{3}{2}$ Vertices}
\vspace{6pt}
\centering
\begin{tabular}{c c c c}
\hline\hline
$\#$ of Derivatives~~~&~~~Vertex~~~~~&~~~~~Nature~~~~~&Exists in\\ [0.5ex]
\hline \vspace{3pt}
1 & $\bar{\psi}_\mu F^{+\mu\nu}\psi_\nu$ & Non-abelian & $D\geq4$\\ \vspace{3pt}
2 & $\left(\bar{\Psi}_{\mu\nu}\,\gamma^{\mu\nu\alpha\beta\lambda}\,\Psi_{\alpha\beta}\right)A_\lambda$ & Abelian & $D\geq5$\\
3 & $\bar{\Psi}_{\mu\alpha}\Psi^\alpha_{~\nu}F^{\mu\nu}$ & Abelian & $D\geq4$\\
\hline\hline\vspace{6pt}
\end{tabular}
\end{table}

Here we parenthetically comment about the nature of the abelian vertices. As it turned out, the vertices that do not deform
the gauge algebra do not deform the gauge transformations either. In other words, if $a_2$ is trivial, so is $a_1$. This is
not accidental at all. In fact, in Section~\ref{sec:noa1} we are going to show that, for a massless particle of arbitrary spin
$s=n+\tfrac{1}{2}$ coupled to a $U(1)$ vector field, the cubic couplings that do not deform the gauge algebra actually do not
deform the gauge transformations and hence only deform the Lagrangian.

\section{Massless Spin 5/2 Coupled to EM}\label{sec:fivehalf}

Now we move on to constructing parity-preserving off-shell cubic vertices for a spin-$\tfrac{5}{2}$ gauge field,
which is a symmetric rank-2 tensor-spinor $\psi_{\mu\nu}$. The original free action is
\beq S^{(0)}[A_\mu,\psi_{\mu\nu}]=\int d^Dx\left[-\tfrac{1}{4}
F_{\mu\nu}^2-\tfrac{1}{2}\left(\bar{\psi}_{\mu\nu}\mathcal{R}^{\mu\nu}-\bar{\mathcal R}^{\mu\nu}\psi_{\mu\nu}
\right)\right],\eeq{5half1} where the tensor $\mathcal{R}^{\mu\nu}$ is related to the spin-$\tfrac{5}{2}$ Fronsdal
tensor, $\mathcal{S}^{\mu\nu}$, as follows. \beq \mathcal{R}^{\mu\nu}=\mathcal{S}^{\mu\nu}-\gamma^{(\mu}\displaystyle
\not{\!\mathcal{S}}^{\nu)}-\tfrac{1}{2}\eta^{\mu\nu}\mathcal{S}',\qquad \mathcal{S}'\equiv\mathcal{S}^\mu_\mu.
\eeq{5half2} Here the photon gauge invariance is as usual, while the fermionic part is gauge invariant under a
\emph{constrained} vector-spinor gauge parameter, $\varepsilon_\mu$, \beq \delta_\varepsilon\psi_{\mu\nu}
=2\partial_{(\mu}\varepsilon_{\nu)},\qquad\displaystyle{\not{\!\varepsilon}}=0.\eeq{5half3} Then, the corresponding
Grassmann-even fermionic ghost, $\xi_\mu$, must also be $\gamma$-traceless: \beq \displaystyle{\not{\!\xi}}=0,
\eeq{5half4} and so will be its antighost. The set of fields and antifields under study are given below.
\beq \Phi^A=\{A_\mu, C, \psi_{\mu\nu}, \xi_\mu\},\qquad \Phi^*_{A}=\{A^{*\mu}, C^*, \bar{\psi}^{*\mu\nu},\bar{\xi}
^{*\mu}\}.\eeq{5half5} The \emph{free} master action, $S_0$, takes the form \beq S_0=\int d^Dx\left[-\tfrac{1}{4}
F_{\mu\nu}^2-\tfrac{1}{2}\left(\bar{\psi}_{\mu\nu}\mathcal{R}^{\mu\nu}-\bar{\mathcal R}^{\mu\nu}\psi_{\mu\nu}\right)
+A^{*\mu}\partial_\mu C - 2(\bar{\psi}^{*\mu\nu}\partial_\mu\xi_\nu-\partial_\mu\bar{\xi}_\nu\psi^{*\mu\nu})\right].
\eeq{5half6} Properties of the various fields and antifields are given in Table 3. Note that the spin-$\tfrac{5}{2}$
curvature tensor is the 2-curl (see  Appendix~\ref{sec:Curvature} for its properties),
\beq \Psi_{\mu_1\nu_1|\mu_2\nu_2}=\left[\partial_{\mu_1}\partial_{\mu_2}\psi_{\nu_1\nu_2}-(\mu_1\leftrightarrow\nu_1)
\right]-(\mu_2\leftrightarrow\nu_2).\eeq{5half7} The cohomology of $\Gamma$ is isomorphic to the space of functions
of (see Appendix~\ref{sec:Gamma1})
\begin{itemize}
 \item The undifferentiated ghosts $\left\{C, \xi_\mu\right\}$, and the $\gamma$-traceless part of the
 1-curl of the spinorial ghost $\xi^{(1)}_{\mu\nu}=2\partial_{[\mu}\xi_{\nu]}$,
 \item The antifields  $\left\{A^{*\mu}, C^*, \bar{\psi}^{*\mu\nu}, \bar{\xi}^{*\mu}\right\}$,
 and their derivatives,
 \item The curvatures $\left\{F_{\mu\nu}, \Psi_{\mu_1\nu_1|\mu_2\nu_2}\right\}$, and their derivatives,
 \item The Fronsdal tensor $\mathcal S_{\mu\nu}$, and its symmetrized derivatives.
\end{itemize}

\begin{table}[ht]
\caption{Properties of the Various Fields \& Antifields ($n=2$)}
\vspace{6pt}
\centering
\begin{tabular}{c c c c c c c}
\hline\hline
$Z$ &$\Gamma(Z)$~~~&~~~$\Delta(Z)$~~~&$pgh(Z)$ &$agh(Z)$ &$gh(Z)$ &$\epsilon(Z)$\\ [0.5ex]
\hline
$A_\mu$ & $\partial_\mu C$ & 0 & 0 & 0 & 0 & 0\\
$C$ & 0 & 0 & 1 & 0 & 1 & 1\\
$A^{*\mu}$ & 0 & $-\partial_\nu F^{\mu\nu}$ & 0 & 1 & $-1$ & 1\\
$C^*$ & 0 & $-\partial_\mu A^{*\mu}$ & 0 & 2 & $-2$ & 0\\ \hline
$\psi_{\mu\nu}$ & $2\partial_{(\mu}\xi_{\nu)}$ & 0 & 0 & 0 & 0 & 1\\
$\xi_\mu$ & 0 & 0 & 1 & 0 & 1 & 0\\
$\bar{\psi}^{*\mu\nu}$ & 0 & $\bar{\mathcal R}^{\mu\nu}$ & 0 & 1 & $-1$ & 0\\
$\bar{\xi}^{*\mu}$ & 0 & $2\partial_\nu\bar{\psi}^{*\mu\nu}$ & 0 & 2 & $-2$ & 1\\
\hline\hline
\end{tabular}
\end{table}
\vspace{6pt}

\subsection{Non-Abelian Vertices}

The set of all possible non-trivial $a_2$'s falls into two subsets: Subset-1 contains \emph{both} the bosonic ghost $C$
and the fermionic ghost $\xi_\mu$, while Subset-2 contains only $\xi_\mu$ but \emph{not} $C$.
\begin{itemize}
 \item Subset-1 $=\left\{\,C\left(\bar{\xi}^*_\mu\,\xi^\mu+\bar{\xi}_\mu\,\xi^{*\mu}\right),
     ~C\left(\bar{\xi}^{*(1)}_{\mu\nu}\,\xi^{(1)\mu\nu}+\bar{\xi}^{\,(1)}_{\mu\nu}\,\xi^{*(1)\mu\nu}\right)\right\}$,
 \item Subset-2 $=\left\{\,C^*\bar{\xi}_\mu\,\xi^\mu,~C^*\bar{\xi}^{\,(1)}_{\mu\nu}\,\xi^{(1)\mu\nu}\right\}$.
\end{itemize}
One can easily verify that other possible rearrangements of derivatives or other possible contractions of the indices, e.g.,
by $\gamma$-matrices, all give trivial terms, thanks to the $\gamma$-tracelessness of the fermionic ghost and its antighost.
Here, the term $C\,\bar{\xi}^*_\mu\,\xi^\mu$ corresponds to potential minimal coupling, while the other candidate $a_2$'s
to multipole interactions.

To see which of the $a_2$'s can be lifted to $a_1$, let us solve Eqn.~(\ref{cocycle2}). A computation, similar
to what leads one from Eqn.~(\ref{rs6}) to Eqn.~(\ref{rs8}), shows that both the elements in Subset-1 enjoy such a lift,
thanks to the relations~(\ref{exc})--(\ref{exc1}) among others. Explicitly, \beq a_2=\begin{cases} C\,\bar{\xi}^*_\mu\,\xi^\mu\\ C\,\bar{\xi}^{*(1)}_{\mu\nu}\,\xi^{(1)\mu\nu}\end{cases}\longrightarrow~~a_1=\begin{cases} -\bar{\psi}^{*\mu\nu}
\left(\psi_{\mu\nu}C+2\xi_\mu A_\nu\right)+\tilde{a}_1\\ -\bar{\psi}^{*(1)\mu\nu\|\,\rho}
\left(\psi^{(1)}_{\mu\nu\|\,\rho}C+2\xi^{(1)}_{\mu\nu}A_\rho\right)+\tilde{a}_1,\end{cases}\eeq{5half9} and similarly
for the hermitian conjugate terms. Here $\tilde{a}_1$ is the usual ambiguity: $\Gamma\tilde{a}_1=0$.
To see whether these could further be
lifted to $a_0$'s, we write \beq \Delta a_1=\begin{cases}-\bar{\mathcal R}^{\mu\nu}\left(\psi_{\mu\nu}C+2\xi_\mu A_\nu\right)
+\Delta\tilde{a}_1\\-\bar{\mathcal R}^{(1)\mu\nu\|\,\rho}\left(\psi^{(1)}_{\mu\nu\|\,\rho}C+2\xi^{(1)}_{\mu\nu}A_\rho\right)
+\Delta\tilde{a}_1.\end{cases}\eeq{5half10} It is important to notice that, up to total derivatives, the $\Delta a_1$'s have
an expansion in the basis of \emph{undifferentiated} ghosts, $\omega_I=\left\{C,\,\xi_\mu\right\}$. Because $\Gamma(\Delta\tilde{a}_1)=-\Delta(\Gamma\tilde{a}_1)=0$, the coefficients $\alpha^I$ in the expansion of the
ambiguity will be $\Gamma$-cocycles, i.e., they will be ``invariant polynomials". Clearly, this is not
the case for the unambiguous pieces; in fact, their expansion coefficients $\beta^I$ are \emph{not} even cocycles of
$H^0(\Gamma|d)$\,\footnote{But still, because of Eqn.~(\ref{rs8.5}), one must have $[\beta^I\omega_I]\in H^1(\Gamma|d)$,
and indeed this is the case.}. Schematically, \beq \Delta a_1=\left(\alpha^I+\beta^I\right)\omega_I+d(...);\qquad
\Gamma\alpha^I=0,\qquad \Gamma\beta^I\not=d(...).\eeq{5half11} Now, $\Gamma a_0$ is a $pgh$-1 object that can be expanded,
up to a total derivative, in the basis of $\left\{\partial_\mu C, \partial_{(\mu}\xi_{\nu)}\right\}$. Then, obviously, one
can also expand it in the undifferentiated ghosts $\omega_I$: \beq \Gamma a_0=-\left(\partial\cdot J\right)^I\omega_I+d(...).
\eeq{correction1} One can plug the respective expansions~(\ref{5half11}) and~(\ref{correction1}) for $\Delta a_1$ and
$\Gamma a_0$ into the consistency condition~(\ref{cocycle3}), and then take a functional derivative w.r.t.
$\omega_I=\left\{C,\,\xi_\mu\right\}$ to find that \beq \alpha^I+\beta^I=\partial\cdot J^I=d(...).\eeq{5half12} But if
this is true, then $\Gamma\left(\alpha^I+
\beta^I\right)=d(...)$, which is in direct contradiction with the properties of $\alpha^I$ and $\beta^I$, given
in~(\ref{5half11})\,\footnote{For the would-be minimal coupling, the impossibility can also be seen as a consequence of
$\alpha^I$ containing too many derivatives compared to $\beta^I$. We have used this argument for spin $\tfrac{3}{2}$.}.
The conclusion is that none of the $a_2$'s in Subset-1 can be lifted all the way to $a_0$. It is important to notice
that this obstruction originates from the very nature of the $a_2$'s themselves, namely each of them contains
\emph{both} the ghosts.

For Subset-2, the analysis simplifies because \emph{only} one term, $C^*\bar{\xi}^{\,(1)}_{\mu\nu}\,\xi^{(1)\mu\nu}$,
with the maximum number of derivatives, can be lifted to an $a_1$. For the other term we have
\bea \Delta\left(C^*\bar{\xi}_{\nu}\,\xi^\nu\right)=A^{*\mu}\left(\bar{\xi}^\nu\partial_\mu\xi_\nu+\partial_\mu\bar{\xi}
_\nu\,\xi^\nu\right)+d(...)\eea{5half13} Because one can write $\partial_\mu\xi_\nu=\partial_{[\mu}\xi_{\nu]}+
\partial_{(\mu}\xi_{\nu)}$, which is the sum of a non-trivial and a trivial element in the cohomology of $\Gamma$, the
right hand side of Eqn.~(\ref{5half13}) cannot be $\Gamma$-exact modulo $d$. Therefore, the candidate $C^*\bar{\xi}_{\mu}
\,\xi^\mu$ is ruled out. However, one finds that \bea \Delta\left(C^*\bar{\xi}^{\,(1)}_{\mu\nu}\,\xi^{(1)\mu\nu}\right)
&=&A^{*\rho}\left(\bar{\xi}^{\,(1)\mu\nu}\partial_\rho\xi^{(1)}_{\mu\nu}+\partial_\rho\bar{\xi}^{\,(1)}_{\mu\nu}\,
\xi^{(1)\mu\nu}\right)+d(...)\nonumber\\&=&\Gamma\left[A^{*\rho}\left(\bar{\psi}^{\,(1)}_{\mu\nu\|\,\rho}\,\xi^{(1)\mu\nu}
-\bar{\xi}^{\,(1)\mu\nu}\,\psi^{(1)}_{\mu\nu\|\,\rho}\right)\right]+d(...),\eea{5half14} thanks to the
relation~(\ref{exc1}). Thus, indeed, $C^*\bar{\xi}^{\,(1)}_{\mu\nu}\,\xi^{(1)\mu\nu}$ gets lifted to an $a_1$:
\beq a_2=C^*\bar{\xi}^{\,(1)}_{\mu\nu}\,\xi^{(1)\mu\nu}\longrightarrow~~a_1=-A^{*\rho}\left(\bar{\psi}^{\,(1)}_
{\mu\nu\|\,\rho}\,\xi^{(1)\mu\nu}-\bar{\xi}^{\,(1)\mu\nu}\,\psi^{(1)}_{\mu\nu\|\,\rho}\right)+\tilde{a}_1.\eeq{5half15}
To see if this $a_1$ can be lifted to an $a_0$, we compute its $\Delta$ variation, \beq \Delta a_1=\Gamma\left(\bar{\psi}
^{\,(1)}_{\alpha\beta\|\,\mu}\,F^{\mu\nu}\psi^{(1)\alpha\beta\|}_{~~~~~~~\nu}\right)+\tfrac{1}{2}F^{\mu\nu}\left(
\bar{\Psi}_{\mu\nu|\alpha\beta}\,\xi^{(1)\alpha\beta}-\bar{\xi}^{\,(1)\alpha\beta}\Psi_{\mu\nu|\alpha\beta}\right)
+\Delta\tilde{a}_1+d(...).\eeq{5half16} This equation bears striking resemblance with its spin-$\tfrac{3}{2}$ counterpart
Eqn.~(\ref{rs9}). We recall that, in the latter, cancelation of non-$\Gamma$-exact terms was possible by the insertion of
identity~(\ref{id0}) in the contraction of curvatures, the Bianchi identity $\partial_{[\mu}F_{\nu\rho]}=0$, and the fermion
EoMs in terms of curvature, $\gamma^\mu\Psi_{\mu\nu}=0, \gamma^{\mu\nu}\Psi_{\mu\nu}=0$. In the present case as well, as
shown in Appendix~\ref{sec:Curvature}, the fermion EoMs can be written as the $\gamma$-traces of the curvature,
$\gamma^\mu\Psi_{\mu\nu|\alpha\beta}=0, \gamma^{\mu\nu}\Psi_{\mu\nu|\alpha\beta}=0$. Therefore, the non-$\Gamma$-exact
terms from the unambiguous piece in~(\ref{5half16}) can indeed be canceled by the $\Delta$
variation of a $\Gamma$-closed ambiguity, \beq \Delta\tilde{a}_1=-\tfrac{1}{4}\left(\displaystyle{\not{\!\bar{\Psi}}}
_{\alpha\beta}\displaystyle{\not{\!F}}-4\bar{\Psi}_{\mu\nu|\alpha\beta}\gamma^{\mu}F^{\nu\rho}\gamma_\rho\right)\xi^{(1)
\alpha\beta}+\text{h.c.}\eeq{5half17} Thus, we have a lift all the way to $a_0$, the latter being a 3-derivative non-abelian
vertex \beq a_0=-\bar{\psi}^{\,(1)}_{\alpha\beta\|\,\mu}\,F^{+\mu\nu}\psi^{(1)\alpha\beta\|}_{~~~~~~~\nu}\,.\eeq{5half18}

\subsection{Abelian Vertices}\label{sec:a1a05/2}

In this case, all the statements~(\ref{rs17})--(\ref{rs18}) hold, and the current in the vertex, $a_0=j^\mu A_\mu$,
is an invariant polynomial, which takes the most general form \beq j^\lambda=\bar{\Psi}_{\mu_1\nu_1|\mu_2\nu_2}\,
X^{\mu_1\nu_1\alpha_1\beta_1\lambda\mu_2\nu_2\alpha_2\beta_2}\,\Psi_{\alpha_1\beta_1|\alpha_2\beta_2}.\eeq{5half19}
Notice that the Fronsdal tensor, although allowed in principle, cannot appear in the current simply because it would
render the vertex $\Delta$-exact. In view of the spin-$\tfrac{5}{2}$ EoMs and the symmetry properties of the field
strength, one can show, like in Section~\ref{sec:a1a0}, that
$X^{\mu_1\nu_1\alpha_1\beta_1\lambda\mu_2\nu_2\alpha_2\beta_2}$ can contain at most one derivative, which must carry
one of the indices.

When $X$ does not contain any derivative, the corresponding vertex will contain 4. In this case, we have the candidate $X^{\mu_1\nu_1\alpha_1\beta_1\lambda\mu_2\nu_2\alpha_2\beta_2}=-2\eta^{\mu_1\nu_1|\alpha_1\beta_1}\eta^{\mu_2\nu_2|
\alpha_2\beta_2}\gamma^\lambda$. But again, the identities~(\ref{id0}) and~(\ref{id3}) tell us that the resulting
vertex deforms nothing: \beq a_0\approx\left(\bar{\Psi}_{\mu_1\nu_1|\mu_2\nu_2}\gamma^{\mu_1\nu_1\alpha_1\beta_1
\lambda}\Psi_{\alpha_1\beta_1|}^{~~~~~\,\mu_2\nu_2}\right)A_\lambda.\eeq{5half22}

Finally, the 1-derivative candidate is $X^{\mu_1\nu_1\alpha_1\beta_1\lambda\mu_2\nu_2\alpha_2\beta_2}=
\tfrac{1}{2}\eta^{\mu_1\nu_1|\alpha_1\beta_1}\eta^{\mu_2\nu_2|\alpha_2\beta_2}\overset\leftrightarrow\partial\,^\lambda$,
which is equivalent to a 5-derivative 3-curvature term (Born-Infeld type),
\beq a_0\approx\bar{\Psi}_{\mu_1\nu_1|\mu_2\rho}\Psi^{\mu_1\nu_1|\rho}_{~~~~~~\nu_2}F^{\mu_2\nu_2}.\eeq{5half25}
Below we present a summary table for all possible $1-\tfrac{5}{2}-\tfrac{5}{2}$ vertices.
\begin{table}[ht]
\caption{Summary of $1-\tfrac{5}{2}-\tfrac{5}{2}$ Vertices}
\vspace{6pt}
\centering
\begin{tabular}{c c c c}
\hline\hline
$\#$ of Derivatives~~~&~~~Vertex~~~~~&~~~~~Nature~~~~~&Exists in\\ [0.5ex]
\hline \vspace{5pt}
3 & $\bar{\psi}^{(1)}_{\alpha\beta\|\,\mu} F^{+\mu\nu}\psi^{(1)\alpha\beta\|}_{~~~~~~~\nu}$ & Non-abelian & $D\geq4$\\
\vspace{5pt}
4 & $\left(\bar{\Psi}_{\mu\nu|\rho\sigma}\,\gamma^{\mu\nu\alpha\beta\lambda}\,\Psi_{\alpha\beta|}^{~~~\,\rho\sigma}\right)
A_\lambda$ & Abelian & $D\geq5$\\ \vspace{2pt}
5 & $\bar{\Psi}_{\alpha\beta|\mu\rho}\Psi^{\alpha\beta|\rho}_{~~~~\,\nu}F^{\mu\nu}$ & Abelian & $D\geq4$\\
\hline\hline\vspace{6pt}
\end{tabular}
\end{table}

\section{Arbitrary Spin: $\textbf{\textit s}\,\bf{=}\,\textbf{\textit n}+\bf{\tfrac{1}{2}}$}\label{sec:arbitrary}

The set of fields and antifields in this case are \beq \Phi^A=\{A_\mu, C, \psi_{\mu_1...\mu_n}, \xi_{\mu_1...
\mu_{n-1}}\},\qquad \Phi^*_{A}=\{A^{*\mu}, C^*, \bar{\psi}^{*\mu_1...\mu_n}, \bar{\xi}^{*\mu_1...\mu_{n-1}}\}.
\eeq{arb1} For $n>2$, there is an additional constraint that the field-antifield are triply $\gamma$-traceless:
\beq \displaystyle{\not{\!\psi}}^{\prime}_{\mu_1\mu_3...\mu_{n-3}}=0,\qquad \displaystyle{\not{\!\bar{\psi}}}
^{*\prime}_{\mu_1\mu_3...\mu_{n-3}}=0,\eeq{arb2} where prime denotes trace w.r.t Minkowski metric. Besides, the
rank-($n-1$) fermionic ghost and its antighost are $\gamma$-traceless: \beq \displaystyle{\not{\!\xi}}_{\mu_1...
\mu_{n-2}}=0,\qquad \displaystyle{\not{\!\bar{\xi}}}^{\,*}_{\mu_1...\mu_{n-2}}=0.\eeq{arb3}
Properties of the various fields and antifields are given in Table 5.

\begin{table}[ht]
\caption{Properties of the Various Fields \& Antifields ($n=\text{arbitrary}$)}
\vspace{6pt}
\centering
\begin{tabular}{c c c c c c c}
\hline\hline
$Z$ &$\Gamma(Z)$~~~&~~~$\Delta(Z)$~~~&$pgh(Z)$ &$agh(Z)$ &$gh(Z)$ &$\epsilon(Z)$\\ [0.5ex]
\hline
$A_\mu$ & $\partial_\mu C$ & 0 & 0 & 0 & 0 & 0\\
$C$ & 0 & 0 & 1 & 0 & 1 & 1\\
$A^{*\mu}$ & 0 & $-\partial_\nu F^{\mu\nu}$ & 0 & 1 & $-1$ & 1\\
$C^*$ & 0 & $-\partial_\mu A^{*\mu}$ & 0 & 2 & $-2$ & 0\\ \hline
$\psi_{\mu_1...\mu_n}$ & $n\partial_{(\mu_1}\xi_{\mu_2...\mu_n)}$ & 0 & 0 & 0 & 0 & 1\\
$\xi_{\mu_1...\mu_{n-1}}$ & 0 & 0 & 1 & 0 & 1 & 0\\
$\bar{\psi}^{*\mu_1...\mu_n}$ & 0 & $\bar{\mathcal R}^{\mu_1...\mu_n}$ & 0 & 1 & $-1$ & 0\\
$\bar{\xi}^{*\mu_1...\mu_{n-1}}$ & 0 & $n\partial_{\mu_n}\bar{\psi}^{*\mu_1...\mu_n}$ & 0 & 2 & $-2$ & 1\\
\hline\hline
\end{tabular}
\end{table}
\vspace{6pt}

The rank-$n$ tensor-spinor $\mathcal R_{\mu_1...\mu_n}$ appearing in the spin-$s$ EoMs is an arbitrary-spin generalization
of~(\ref{5half2}); it is related to the Fronsdal tensor as \beq \mathcal R_{\mu_1...\mu_n}=\mathcal S_{\mu_1...\mu_n}
-\tfrac{1}{2}n\,\gamma_{(\mu_1}\displaystyle\not{\!\mathcal S}_{\mu_2...\mu_n)}-\tfrac{1}{4}n(n-1)\,\eta_{(\mu_1\mu_2}
\mathcal S^\prime_{\mu_3...\mu_n)}.\eeq{arb4}

The cohomology of $\Gamma$ has already been given in Section~\ref{sec:Gamma}, with the details appearing in
Appendix~\ref{sec:Gamma1}. One can immediately write down the set of all possible non-trivial $a_2$'s. Again, they fall
into two subsets: Subset-1 contains \emph{both} the bosonic ghost $C$ and the fermionic ghost $\xi_{\mu_1...\mu_{n-1}}$,
while Subset-2 contains only $\xi_{\mu_1...\mu_{n-1}}$ but \emph{not} $C$.
\begin{itemize}
 \item Subset-1 $=\left\{\,C\,\bar{\xi}^{*(m)}_{\mu_1\nu_1|...|\mu_m\nu_m\|\nu_{m+1}...\nu_{n-1}}\,
     \xi^{(m)\mu_1\nu_1|...|\mu_m\nu_m\|\nu_{m+1}...\nu_{n-1}}+\text{h.c.}\right\}$,
 \item Subset-2 $=\left\{\,C^*\bar{\xi}^{\,(m)}_{\mu_1\nu_1|...|\mu_m\nu_m\|\nu_{m+1}...\nu_{n-1}}\,
     \xi^{(m)\mu_1\nu_1|...|\mu_m\nu_m\|\nu_{m+1}...\nu_{n-1}}\right\}$.
\end{itemize}
Here, $0\leq m\leq n-1$. As a straightforward generalization of the spin-$\tfrac{5}{2}$ case, one finds that each
element in Subset-1 gets lifted to $a_1$: \bea a_1&=&-\bar{\psi}^{*(m)\mu_1\nu_1|...|\mu_m\nu_m\|\,\nu_{m+1}...\nu_n}\,
\psi^{(m)}_{\mu_1\nu_1|...|\mu_m\nu_m\|\,\nu_{m+1}...\nu_n}C\nonumber\\&&-n\,\bar{\psi}^{*(m)\mu_1\nu_1|...|\mu_m\nu_m\|
\,\nu_{m+1}...\nu_n}\,\xi^{(m)}_{\mu_1\nu_1|...|\mu_m\nu_m\|\,(\nu_{m+1}...\nu_{n-1}}A_{\nu_n)}+\text{h.c.}+\tilde{a}_1.
\eea{arb5} Now, one can compute $\Delta a_1$ and expand it in the basis of $pgh$-1 objects in the cohomology of $\Gamma$,
namely $\omega_I=\left\{C,\,\xi^{(m)}_{\mu_1\nu_1|...|\mu_m\nu_m\|\,\nu_{m+1}...\nu_{n-1}}~|~0\leq m\leq n-1\right\}$.
Upon comparing the expansion coefficients for the unambiguous piece and the ambiguity $\tilde a_1$, again one
can conclude that none of these $a_1$'s can be lifted to an $a_0$. On the other hand, for the elements of Subset-2,
one notices that \bea &&\Delta\left(C^*\bar{\xi}^{\,(m)}_{\mu_1\nu_1|...|\mu_m\nu_m\|\nu_{m+1}...\nu_{n-1}}\,
\xi^{(m)\mu_1\nu_1|...|\mu_m\nu_m\|\nu_{m+1}...\nu_{n-1}}\right)\nonumber\\&&=A^{*\nu_n}\bar{\xi}^{\,(m)\mu_1\nu_1
|...|\mu_m\nu_m\|\nu_{m+1}...\nu_{n-1}}\partial_{\nu_n}\xi^{(m)\mu_1\nu_1|...|\mu_m\nu_m\|}_{~~~~~~~~~~~~~~~~~~\,
\nu_{m+1}...\nu_{n-1}}+\text{h.c.}+d(...).~~~~~~~~~~\eea{arb6} Then, in view of Eqn.~(\ref{exc})--(\ref{exc1}),
it is clear that the right side of the above equation is $\Gamma$-exact modulo $d$ \emph{only} for $m=n-1$. This
rules out, in particular, the would-be minimal coupling corresponding to $m=0$. Therefore, one is left with the lift:
\beq a_1=-A^{*\nu_n}\,\bar{\psi}^{\,(n-1)}_{\mu_1\nu_1|...|\mu_{n-1}\nu_{n-1}\|\,\nu_n}\,\xi^{(n-1)\mu_1\nu_1|...|
\mu_{n-1}\nu_{n-1}}+\text{h.c.}+\tilde{a}_1,\eeq{arb7} whose $\Delta$-variation is given by \bea \Delta a_1&=&
\Gamma\left(\bar{\psi}^{(n-1)}_{\mu_1\nu_1|...|\mu_{n-1}\nu_{n-1}\|\,\mu_n}F^{\mu_n}_{~~\nu_n}\,\psi^{(n-1)\mu_1
\nu_1|...|\mu_{n-1}\nu_{n-1}\|\,\nu_n}\right)+\Delta\tilde{a}_1+d(...)\nonumber\\&&+\tfrac{1}{2}F^{\mu_n\nu_n}
\left(\bar{\Psi}_{\mu_1\nu_1|...|\mu_n\nu_n}\,\xi^{(n-1)\mu_1\nu_1|...|\mu_{n-1}\nu_{n-1}}+\text{h.c.}\right).\eea{arb8}
In view of Eqn.~(\ref{rs9}) and~(\ref{5half16}), pertaining respectively to the spin-$\tfrac{3}{2}$ and spin-$\tfrac{5}{2}$
cases, and the subsequent steps, we realize that it is possible to cancel the non-$\Gamma$-exact terms in~(\ref{arb8})
by inserting identity~(\ref{id0}) in the contraction of curvatures, thanks to the Bianchi identity $\partial_{[\mu}
F_{\nu\rho]}=0$, and to the fermion EoMs in terms of curvature (see Appendix~\ref{sec:Curvature}), $\gamma^{\mu_1}
\Psi_{\mu_1\nu_1|...|\mu_n\nu_n}=0,\,\gamma^{\mu_1\nu_1}\Psi_{\mu_1\nu_1|...|\mu_n\nu_n}=0$. In other words,
$\Delta a_1$ is rendered $\Gamma$-exact modulo $d$ by an appropriate choice of the ambiguity $\tilde a_1$, so that
one finally has \beq a_0=-\bar{\psi}^{(n-1)}_{\mu_1\nu_1|...|\mu_{n-1}\nu_{n-1}\|\,\mu_n}F^{+\mu_n}_{~~~~\nu_n}
\psi^{(n-1)\mu_1\nu_1|...|\mu_{n-1}\nu_{n-1}\|\,\nu_n}.\eeq{arb9} This $(2n-1)$-derivative non-abelian vertex
contains the $(n-1)$-curl of the fermionic field.

For an abelian vertex, $a_0=j^\mu A_\mu$, the gauge-invariant current does not contain the Fronsdal tensor nor
its derivatives, since their presence would render the vertex $\Delta$-exact. Again, non-triviality of the abelian
deformation allows only two possible values for the number of derivatives in the vertex: $2n$ and $2n+1$. The
off-shell vertices can be obtained exactly the same way as for spins $\tfrac{3}{2}$ and $\tfrac{5}{2}$, considered
in Sections~\ref{sec:a1a0} and~\ref{sec:a1a05/2} respectively. A summary table for all possible $1-s-s$ vertices
is given below.

\begin{table}[ht]
\caption{Summary of $1-s-s$ Vertices with $p$ Derivatives}
\vspace{6pt}
\centering
\begin{tabular}{c c c c}
\hline\hline
$p$&~~Vertex~~&~~Nature~~&Exists in\\ [0.5ex]
\hline \vspace{8pt}
$2n-1$ & $\bar{\psi}^{(n-1)}_{\mu_1\nu_1|...|\mu_{n-1}\nu_{n-1}\|\,\mu_n}F^{+\mu_n}_{~~~~\nu_n}\psi^{(n-1)\mu_1\nu_1
|...|\mu_{n-1}\nu_{n-1}\|\,\nu_n}$ & Non-abelian & $D\geq4$\\ \vspace{8pt}
$2n$ & $(\bar{\Psi}_{\mu_1\nu_1|\mu_2\nu_2|...|\mu_n\nu_n}\gamma^{\mu_1\nu_1\alpha_1\beta_1\lambda}\Psi_{\alpha_1\beta_1|}
 ^{~~~~~~\mu_2\nu_2|...|\mu_n\nu_n})A_\lambda$ & Abelian & $D\geq5$\\ \vspace{3pt}
$2n+1$ & $\bar{\Psi}_{\mu_1\nu_1|\mu_2\nu_2|...|\mu_n\alpha}\Psi^{\mu_1\nu_1|\mu_2\nu_2|...|\alpha\nu_n}F^{\mu_n}_{~~\,\nu_n}$
& Abelian & $D\geq4$\\
\hline\hline
\end{tabular}
\end{table}

\section{Abelian Vertices Preserve Gauge Symmetries}\label{sec:noa1}

Abelian vertices are those that do not deform the gauge algebra, i.e., they can only have a trivial $a_2$.
For such a vertex, it is always possible to choose $a_1$ to be $\Gamma\text{-closed}$~\cite{BBH}:
\beq \Gamma a_1=0.\eeq{noa10} If this gets lifted to an $a_0$, one has the cocycle condition~(\ref{cocycle3}),
\beq \Delta a_1+\Gamma a_0+db_0=0.\eeq{noa10.1}

For the $1-s-s$ vertices under study, one can always write a vertex as the photon field $A_\mu$ contracted with a
current $j^\mu$, which is a fermion bilinear: \beq a_0=j^\mu A_\mu.\eeq{noa10.2} One can always choose the current
such that it satisfies \beq \Gamma j^\mu=0,\qquad \partial_\mu j^\mu=\Delta M,\qquad \Gamma M=0.\eeq{noa1.3}
To see this, let us note that the $a_1$ corresponding to~(\ref{noa10.2}) has the general form
\beq a_1=MC+\left(\bar{P}_{\mu_1...\mu_{n-1}}\xi^{\mu_1...\mu_{n-1}}-\bar{\xi}_{\mu_1...\mu_{n-1}}P^{\mu_1...\mu_{n-1}}
\right)+a_1',\eeq{new1} where $M$ and $P_{\mu_1...\mu_{n-1}}$ belong to $H(\Gamma)$, with $pgh=0$, $agh=1$,
and $a_1'$ stands for expansion terms in the ghost-curls. Given~(\ref{noa10.2}) and ~(\ref{new1}), the
condition~(\ref{noa10.1}) reads \beq \Gamma\left(j^\mu A_\mu\right)+\Delta MC+\left(\Delta\bar{P}_{\mu_1...\mu_{n-1}}
\xi^{\mu_1...\mu_{n-1}}-\bar{\xi}_{\mu_1...\mu_{n-1}}\Delta P^{\mu_1...\mu_{n-1}}\right)+\Delta a_1'+db_0=0.\eeq{new100}
Now, $P_{\mu_1...\mu_{n-1}}$ consists of two kinds of terms: one contains the antifield $A^{*\mu}$ and its
derivatives, and the other contains the antifield $\psi^{*\nu_1...\nu_n}$ and its derivatives. The former one
also contains (derivatives of) the Fronsdal tensor $\mathcal S_{\nu_1...\nu_n}$, or (derivatives of) the curvature
$\Psi_{\rho_1\nu_1|...|\rho_n\nu_n}$, while the latter one contains (derivatives of) the EM field strength $F_{\mu\nu}$.
One can choose to get rid of derivatives on $A^{*\mu}$ and $F_{\mu\nu}$ by using the Leibniz rule,
\bea P_{\mu_1...\mu_{n-1}}&=&A^{*\mu}\left(\vec{P}_{\mu,\,\mu_1...\mu_{n-1}}^{(\mathcal S)~~~~~~~\,\nu_1...\nu_n}\mathcal S_{\nu_1...\nu_n}+\vec{P}_{\mu,\,\mu_1...\mu_{n-1}}^{(\Psi)~~~~~~~\,\rho_1\nu_1|...|\rho_n\nu_n}\Psi_{\rho_1\nu_1|...|
\rho_n\nu_n}\right)\nonumber\\&&+F^{\mu\nu}\left(\vec{P}_{\mu\nu,\,\mu_1...\mu_{n-1}}^{(\psi^*)~~~~~~~\,\nu_1...\nu_n}
\psi^*_{\nu_1...\nu_n}\right)+\partial^{\mu_n}p_{\mu_1...\mu_n},\eea{new2} where $\Gamma p_{\mu_1...\mu_n}=0$, and the
$\vec P$'s are operators acting to the right. Notice that the quantity in the parentheses in the first line is both
$\Gamma$-closed and $\Delta$-exact\,\footnote{$\Delta$-exactness of the first term is manifest, while in the second,
the presence of the curvature admits only $\Delta$-exact terms, like its own $\gamma$-traces and divergences (see
Appendix~\ref{sec:Curvature}).}. One can take the $\Delta$-variation of~(\ref{new2}), and then add a total derivative in
order to cast it in the form \beq \Delta P_{\mu_1...\mu_{n-1}}=\tfrac{1}{2}F^{\mu\nu}\Delta Q_{[\mu\nu],\,\mu_1...\mu_{n-1}}
+\partial^{\mu_n}\Delta q_{\mu_1...\mu_n},\eeq{new3} where, $\Gamma Q_{[\mu\nu],\,\mu_1...\mu_{n-1}}=0$, $\Gamma q_{\mu_1...
\mu_n}=0$. Therefore, we have \beq \bar{\xi}_{\mu_1...\mu_{n-1}}\Delta P^{\mu_1...\mu_{n-1}}=A_\mu\Delta\left[\partial_\nu
\left(\bar\xi_{\mu_1...\mu_{n-1}}Q^{[\mu\nu],\,\mu_1...\mu_{n-1}}\right)\right]-\bar{\xi}_{\mu_1...\mu_{n-1}}\overset\leftarrow
\partial_{\mu_n}\Delta q^{\mu_1...\mu_n}+d(...).\eeq{new4} The second term on the right side is $\Gamma$-closed,
and can be broken as a $\Gamma$-exact piece plus terms involving the ghost-curls. The latter can always be canceled in the
cocycle condition~(\ref{new100}) by appropriately choosing the similar terms coming from $a_1'$. Thus,
\bea \Gamma\left[j^\mu A_\mu+\Delta\left(\bar\psi_{\mu_1...\mu_n}q^{\mu_1...\mu_n}+\text{h.c.}\right)\right]+\Delta MC
~~~~~~~~~~~~~~~~~~~~~~~~~~~~~~~~~~~~~\nonumber\\-A_\mu\Delta\left[\partial_\nu\left(\bar\xi_{\mu_1...\mu_{n-1}}
Q^{[\mu\nu],\,\mu_1...\mu_{n-1}}\right)+\text{h.c.}\right]+d(...)=0.\eea{new5} The $\Delta$-exact term added to the
original vertex $j^\mu A_\mu$ is trivial, and therefore can be dropped. Now we are left with
\beq A_\mu\left[\Gamma j^\mu-\Delta\left(\partial_\nu\left(\bar\xi_{\mu_1...\mu_{n-1}}Q^{[\mu\nu],\,\mu_1...\mu_{n-1}}\right)
+\text{h.c.}\right)\right]+\left(\Delta M-\partial_\mu j^\mu\right)C+d(...)=0.\eeq{new6} Taking functional derivative
w.r.t. $C$ produces part of the sought-after conditions~(\ref{noa1.3}), \beq \partial_\mu j^\mu=\Delta M,\qquad
\Gamma M=0,\eeq{new7} while the functional derivative w.r.t. $A_\mu$ gives \beq \Gamma j^\mu=\partial_\nu
\left(\bar\xi_{\mu_1...\mu_{n-1}}\Delta Q^{[\mu\nu],\,\mu_1...\mu_{n-1}}\right)+\text{h.c.},\qquad \Gamma Q^{[\mu\nu],
\,\mu_1...\mu_{n-1}}=0.\eeq{new8} The expression for $\Gamma j^\mu$ has to be $\Gamma$-exact. This demands that
$\partial_\nu Q^{[\mu\nu],\,\mu_1...\mu_{n-1}}$ be $\Delta$-closed, and that $Q^{[\mu\nu],\,\mu_1...\mu_{n-1}}$
have the interchange symmetry $\nu\leftrightarrow\mu_i$, $i=1,...,n-1$. Then, \beq j^\alpha=\tilde{j}^\alpha
+\Delta\left(\tfrac{1}{n}\,\bar\psi_{\mu_1...\mu_n}Q^{[\alpha\mu_1],\,\mu_2...\mu_n}+\text{h.c.}\right),
\qquad \Gamma \tilde{j}^\alpha=0.\eeq{new9} Therefore, by field redefinitions, the current can always be made
gauge invariant: \beq \Gamma j^\mu=0.\eeq{new10} This completes the proof of~(\ref{noa1.3}).
Then, from~(\ref{noa10.1}), one obtains the lift: \beq a_1=MC.\eeq{noa12}

Now we will show that $M$ must be $\Delta$-exact modulo $d$. We recall that $M$ belongs to the cohomology of $\Gamma$,
with $pgh=0$, $agh=1$. It will contain (derivatives of) the fermionic antifield, and (derivatives of) the Fronsdal
tensor $\mathcal S_{\nu_1...\nu_n}$ or the curvature $\Psi_{\rho_1\nu_1|...|\rho_n\nu_n}$. However, one can choose
to have no derivatives of the antifield by using the Leibniz rule. Thus $M$ has the most general
form \beq M=\bar{\psi}^{*\mu_1...\mu_n}\left(\vec{M}_{\mu_1...\mu_n}^{(\mathcal S)~~~\,\nu_1...\nu_n}\mathcal S_{\nu_1...\nu_n}+\vec{M}_{\mu_1...\mu_n}^{(\Psi)~~~\,\rho_1\nu_1|...|\rho_n\nu_n}\Psi_{\rho_1\nu_1|...|
\rho_n\nu_n}\right)+\partial^\mu m_\mu-\text{h.c.},\eeq{noa13} where $\Gamma m_\mu=0$, and the operators $\vec M$'s
act to the right. The first term in the parentheses is manifestly $\Delta$-exact, while the second one must contain
either a $\gamma$-trace and or a divergence of the curvature, which are $\Delta$-exact too (see
Appendix~\ref{sec:Curvature}). Therefore, $M$ must be $\Delta$-exact modulo $d$. This means that $a_1$, given
in~(\ref{noa12}), can be rendered trivial by adding a $\Delta$-exact piece in $a_0$~\cite{BBH}, and so the vertex
will be gauge invariant up to a total derivative: \beq \Gamma a_0+db_0=0.\eeq{noa14} In other words, one can always
add a $\Delta$-exact term in $a_0$, so that the new current is identically conserved~\cite{BBH}: \beq j^\mu
\rightarrow j^{\prime\mu}=j^\mu+\Delta k^\mu=\partial_\nu\mathcal A^{\mu\nu}, \qquad \mathcal A^{\mu\nu}=-\mathcal
A^{\nu\mu}.\eeq{noa15} Thus we have proved that no abelian vertex can deform the gauge transformations.

\section{Comparative Study of Vertices}\label{sec:compare}

We have found that the possible number of derivatives in a $1-s-s$ vertex, with $s=n+\tfrac{1}{2}$, is restricted
to the values: $2n-1, 2n$, and $2n+1$. Moreover, the $2n$-derivative vertex exists only in $D\geq 5$. These are in
complete agreement with Metsaev's light-cone-formulation results~\cite{Metsaev}. While the light-cone vertices are
maximally gauge fixed, the corresponding covariant on-shell vertices were also written down in~\cite{Metsaev} for
lower spins, from previously known results. These on-shell vertices are partially gauge fixed, with the gauge choice
being the transverse-traceless gauge (TT gauge), \beq \partial^{\mu_1}\psi_{\mu_1...\mu_n}=0,\quad\gamma^{\mu_1}
\psi_{\mu_1...\mu_n}=0,\qquad \partial^\mu A_\mu=0 .\eeq{TTgauge} Note that in this gauge the fermion and photon
EoMs boil down to \beq \displaystyle{\not{\!\partial\,}}\psi_{\mu_1...\mu_n}=0,\qquad \Box A_\mu=0.\eeq{TTgaugeEoM}

We will find that our off-shell vertices reduce in the TT gauge to the on-shell ones given in~\cite{Metsaev}. So do
the Sagnotti-Taronna (ST) off-shell vertices~\cite{Taronna}, as we will see. If two vertices are shown to match in
a particular gauge, say the TT gauge, the full off-shell ones must be equivalent, i.e., differ only by terms that are
$\Delta$-exact modulo $d$. Still, for the simplest case of spin $\tfrac{3}{2}$, we will make explicit the off-shell
equivalence of the ST vertices with ours. For $s\geq\tfrac{5}{2}$, we match our vertices with the ST ones in the
TT gauge.

The ST off-shell vertices, when read off in the most naive way, contain many terms, and it is not straightforward at
all to see that some of them actually vanish in $D=4$. In comparison, the off-shell vertices we present for arbitrary
spin are rather neat in form, and the absence of some of them in $D=4$ is obvious from inspection.

We will denote a $p$-derivative off-shell vertex of ours as $V^{(p)}$, and its Sagnotti-Taronna counterpart as
$V^{(p)}_{\text{ST}}$. The corresponding TT-gauge vertex will be denoted as $V^{(p)}_{\text{TT}}$\,\footnote{Both
the spins $m+\tfrac{1}{2}$ and $m+\tfrac{3}{2}$ will respectively have one vertex with $2m+1$ derivatives. Our
notation should not cause any confusion, as we will be considering one spin at a time.}.

\subsection{1---3/2---3/2 Vertices}\label{sec:com3/2}

Our 1-derivative off-shell $1-\tfrac{3}{2}-\tfrac{3}{2}$ vertex is given by \beq V^{(1)}=\bar{\psi}_\mu F^{+\mu\nu}
\psi_\nu=\bar{\psi}_\mu\left(\eta^{\mu\nu|\alpha\beta}+\tfrac{1}{2}\gamma^{\mu\nu\alpha\beta}\right)F_{\alpha\beta}
\psi_\nu.\eeq{com1} To see what it reduces to in the TT gauge, let us rewrite identity~(\ref{id0}) as
\bea \eta^{\mu\nu|\alpha\beta}+\tfrac{1}{2}\gamma^{\mu\nu\alpha\beta}&=&-\tfrac{1}{2}\eta^{\mu\nu}\gamma^{\alpha\beta}
+\tfrac{1}{2}\gamma^\mu\gamma^\nu\gamma^{\alpha\beta}-2\gamma^{[\mu}\eta^{\nu][\alpha}\gamma^{\beta]}\nonumber\\
&=&2\left(\eta^{\mu\nu|\alpha\beta}-\tfrac{1}{4}\eta^{\mu\nu}\gamma^{\alpha\beta}\right)+\tfrac{1}{4}\left(\gamma^\mu
\gamma^\nu\gamma^{\alpha\beta}+\gamma^{\alpha\beta}\gamma^\mu\gamma^\nu\right),\eea{id4}
where in the second line we have used $2\gamma^{[\mu}\eta^{\nu][\alpha}\gamma^{\beta]}=\tfrac{1}{4}\left(\gamma^\mu
\gamma^\nu\gamma^{\alpha\beta}-\gamma^{\alpha\beta}\gamma^\mu\gamma^\nu\right)-2\eta^{\mu\nu|\alpha\beta}$. Upon
insertion of identity~(\ref{id4}) into the vertex~(\ref{com1}), one finds \beq V^{(1)}=2\left(\bar{\psi}_\mu F^{\mu\nu}
\psi_\nu-\tfrac{1}{4}\bar{\psi}_\mu\displaystyle{\not{\!F}}\psi^\mu\right)+\tfrac{1}{4}\left(\displaystyle{\not{\!\bar
{\psi}}}\gamma^\mu\displaystyle{\not{\!F}}\psi_\mu+\bar{\psi}_\mu\displaystyle{\not{\!F}}\gamma^\mu\displaystyle{\not
{\!\psi}}\right).\eeq{com2} On the other hand, the 1-derivative ST vertex reads~\cite{Taronna} \beq V^{(1)}_{\text{ST}}
=\bar{\psi}^\mu(\partial_\nu\psi_\mu)A^\nu-(\partial_\mu\bar{\psi}_\nu)\psi^\nu A^\mu+\bar{\psi}_\mu\psi_\nu(\partial
^\mu A^\nu)-\bar{\psi}_\mu(\partial^\mu\psi^\nu)A_\nu+(\partial_\mu\bar{\psi}_\nu)\psi^\mu A^\nu-\bar{\psi}_\mu\psi_\nu
(\partial^\nu A^\mu).\eeq{com3} Integrating by parts the 2nd, 4th and 5th terms on the right hand side, we obtain
\beq V^{(1)}_{\text{ST}}=2\bar{\psi}_\mu F^{\mu\nu}\psi_\nu+2\bar{\psi}_\mu A\cdot\partial\psi^\mu+\bar\psi_\mu
(\partial\cdot A)\psi^\mu+(\partial\cdot\bar\psi)A\cdot\psi-\bar\psi\cdot A(\partial\cdot\psi)+d(...).\eeq{com4}
Let us take the 2nd term on the right hand side and replace $\eta^{\alpha\beta}=\gamma^{(\alpha}\gamma^{\beta)}$ in the
operator $(A\cdot\partial)$. Also in the 3rd term we replace $\eta^{\alpha\beta}=\gamma^{\alpha}\gamma^{\beta}-
\gamma^{\alpha\beta}$ in $(\partial\cdot A)$. The result is \beq 2\bar{\psi}_\mu A\cdot\partial\psi^\mu+\bar\psi_\mu
(\partial\cdot A)\psi^\mu=-\tfrac{1}{2}\bar{\psi}_\mu\displaystyle{\not{\!F}}\psi^\mu+\bar{\psi}_\mu\displaystyle
{\not{\!\!A\,}}(\displaystyle{\not{\!\partial\,}}\psi^\mu)-(\displaystyle{\not{\!\partial\,}}\bar{\psi}_\mu)
\displaystyle{\not{\!\!A\,}}\psi^\mu+d(...),\eeq{com5} which, when plugged into the vertex~(\ref{com4}) gives
\beq V^{(1)}_{\text{ST}}=2\left(\bar{\psi}_\mu F^{\mu\nu}\psi_\nu-\tfrac{1}{4}\bar{\psi}_\mu\displaystyle{\not{\!F}}
\psi^\mu\right)+\left[\bar{\psi}_\mu\displaystyle{\not{\!\!A\,}}(\displaystyle{\not{\!\partial\,}}\psi^\mu)
-\bar{\psi}\cdot A(\partial\cdot\psi)+\text{h.c.}\right]+d(...).\eeq{com6} It is obvious that both the off-shell
vertices~(\ref{com2}) and~(\ref{com6}) reduce in the TT gauge to \beq V^{(1)}_{\text{TT}}=2\left(\bar{\psi}_\mu F^{\mu\nu}\psi_\nu-\tfrac{1}{4}\bar{\psi}_\mu\displaystyle{\not{\!F}}\psi^\mu\right).\eeq{com7} This is precisely
the on-shell 1-derivative vertex reported by Metsaev~\cite{Metsaev}. To see explicitly that the off-shell vertices
are also equivalent, we subtract~(\ref{com6}) from~(\ref{com2}) to get \beq V^{(1)}-V^{(1)}_{\text{ST}}=\tfrac{1}{4}
\left(\displaystyle{\not{\!\bar{\psi}}}\gamma^\mu\displaystyle{\not{\!F}}\psi_\mu+\bar{\psi}_\mu\displaystyle{\not
{\!F}}\gamma^\mu\displaystyle{\not{\!\psi}}\right)-\left[\bar{\psi}_\mu\displaystyle{\not{\!\!A\,}}(\displaystyle
{\not{\!\partial\,}}\psi^\mu)-\bar{\psi}\cdot A(\partial\cdot\psi)+\text{h.c.}\right]+d(...).\eeq{com8} Now we
make use of the identity \beq \left[\gamma^\mu,\gamma^{\alpha\beta}\right]=4\eta^{\mu[\alpha}\gamma^{\beta]},
\eeq{id8} in order to be able to pass $\gamma^\mu$ past $\displaystyle{\not{\!F}}$ in both the terms in the
parentheses on the right hand side of Eqn.~(\ref{com8}). As a result, we will obtain, among others, the term
$\tfrac{1}{2}\displaystyle{\not{\!\bar{\psi}}}\displaystyle{\not{\!F}}\displaystyle{\not{\!\psi}}$, in which we
replace $\displaystyle{\not{\!F}}=\displaystyle{\not{\!\partial\,}}\displaystyle{\not{\!\!A}}-\partial\cdot A$.
Now in all the resulting terms we perform integrations by parts such that no derivative acts on the photon field.
The final result is \beq V^{(1)}-V^{(1)}_{\text{ST}}=\left[2\bar\psi^{[\mu}A^{\nu]}\gamma_\mu\left(\displaystyle
{\not{\!\partial\,}}\psi_\nu-\partial_\nu\displaystyle{\not{\!\psi}}\right)-\bar\psi^{\mu}A^{\nu}\gamma_{\mu\nu}
\left(\partial\cdot\psi-\displaystyle{\not{\!\partial\,}}\displaystyle{\not{\!\psi}}\right)+\text{h.c.}\right]
+d(...).\eeq{com9} This is manifestly $\Delta$-exact modulo $d$, which proves the equivalence of the off-shell
vertices: \beq V^{(1)}\approx V^{(1)}_{\text{ST}}.\eeq{com10}

Next, we consider the 2-derivative vertex, \beq V^{(2)}=\left(\bar{\Psi}_{\mu\nu}\,\gamma^{\mu\nu\alpha\beta\lambda}
\,\Psi_{\alpha\beta}\right)A_\lambda\approx-2\left(\bar{\Psi}_{\mu\nu}\,\gamma^\rho\,\Psi^{\mu\nu}\right)A_\rho.
\eeq{com11} One can use the definition $\Psi_{\mu\nu}=2\partial_{[\mu}\psi_{\nu]}$ to rewrite it as
\beq V^{(2)}\approx-4\bar\psi_\alpha\overset\leftarrow{\partial}_{\mu}\displaystyle{\not{\!\!A}}\,\partial^\mu
\psi^\alpha+2\left(\bar\psi_\alpha\overset\leftarrow{\partial}_{\mu}\displaystyle{\not{\!\!A}}\,\partial^\alpha
\psi^\mu+\text{h.c.}\right).\eeq{com12} In the 1st term, we can use the 3-box rule, already given in
Section~(\ref{sec:a1a0}), \beq 2\partial_\mu X\partial^\mu Y=\Box(XY)-X(\Box Y)-(\Box X)Y,\eeq{3box} and perform
a double integration by parts in order to have a $\Box$ acting on the photon field. In the 2nd term on the right
hand side of~(\ref{com11}) one can integrate by parts w.r.t. any of the derivatives. When the derivative acts
on the photon field, one can use $\partial_\mu A_\nu=F_{\mu\nu}+\partial_\nu A_\mu$ to rewrite it in terms of the
field strength. The result is \bea V^{(2)}&\approx&2\left(\bar\psi_\alpha\gamma^\mu\partial^\alpha\psi^\nu-\bar
\psi^\mu\overset\leftarrow{\partial}\,^\alpha\gamma^\nu\psi_\alpha\right)F_{\mu\nu}-2\left[\left(\bar\psi_\alpha
\gamma^\mu\partial^\alpha\psi^\nu\right)\partial_\mu A_\nu+\text{h.c.}\right]\nonumber\\&&-2\bar\psi_\alpha
\Box\displaystyle{\not{\!\!A}}\,\psi^\alpha+2\left[\bar\psi_\alpha{\not{\!\!A}}\,\left(\Box\psi^\alpha
-\partial^\alpha\partial\cdot\psi\right)+\text{h.c.}\right].\eea{com13} Now, in the last term of the first line we
perform integration by parts so that no derivative acts on the photon field. On the other hand, the last term in
the second line is $\Delta$-exact, and therefore can be dropped. Thus we are left with
\bea V^{(2)}&\approx&2\left(\bar\psi_\alpha\gamma^\mu\partial^\alpha\psi^\nu-\bar\psi^\mu\overset\leftarrow{\partial}
\,^\alpha\gamma^\nu\psi_\alpha\right)F_{\mu\nu}\nonumber\\&&+2\left[\left(\bar\psi_\alpha\partial^\alpha\displaystyle
{\not{\!\partial}}\,\psi^\nu+\bar\psi_\alpha\overset\leftarrow{\displaystyle{\not{\!\partial}}}\,\partial^\alpha
\psi^\nu\right)A_\nu+\text{h.c.}\right]-2\bar\psi_\alpha\Box\displaystyle{\not{\!\!A}}\,\psi^\alpha.\eea{com14}
As one reads off the 2-derivative ST vertex, it gives \bea V^{(2)}_{\text{ST}}&=&-\bar\psi^\mu\gamma^\alpha
(\partial_\mu\psi_\nu)\partial^\nu A_\alpha+\bar\psi^\mu\gamma^\alpha\psi^\nu\partial_\mu\partial_\nu A_\alpha+
\bar\psi^\mu\overset\leftarrow{\partial}\,^\nu\gamma^\alpha(\partial_\mu\psi_\nu)A_\alpha-\bar\psi^\mu\overset
\leftarrow{\partial}\,^\nu\gamma^\alpha\psi_\nu\partial_\mu A_\alpha\nonumber\\&&-(\partial\cdot\bar\psi)
\displaystyle{\not{\!\!A}}\,\partial\cdot\psi-\displaystyle{\not{\!\bar\psi}}\,\psi^\nu\partial_\nu\partial\cdot A
+\displaystyle{\not{\!\bar\psi}}\,\overset\leftarrow{\partial}_\nu\psi^\nu\partial\cdot A+\bar\psi^\mu(\partial_\mu
\displaystyle{\not{\!\psi}})\partial\cdot A\nonumber\\&&-\bar\psi^\mu\displaystyle\not{\!\psi}\,\partial_\mu\partial
\cdot A-\displaystyle{\not{\!\bar\psi}}\,(\partial\cdot\psi)\partial\cdot A-(\partial\cdot\bar\psi)\displaystyle{\not
{\!\psi}}\,\partial\cdot A\,.\eea{com15} As we mentioned already, in this form it is not evident at all that this vertex
vanishes for $D=4$. Let us integrate by parts the 2nd and 3rd terms in the first line of Eqn.~(\ref{com15}), w.r.t.
$\partial_\mu$. The 2nd term in the second line and the 1st term in the third line contain the gradient of
$\partial\cdot A$; we integrate by parts the gradient in both these terms. Thus we have \bea V^{(2)}_{\text{ST}}&=&
-2\bar\psi_\mu\gamma^\alpha(\partial^\mu\psi^\nu)\partial_\nu A_\alpha-(\partial\cdot\bar\psi)(\partial_\nu\displaystyle
{\not{\!\!A}})\psi^\nu-2\bar\psi^\mu\overset\leftarrow{\partial}\,^\nu\gamma^\alpha\psi_\nu\partial_\mu A_\alpha
-(\partial\cdot\bar\psi)\overset\leftarrow\partial_\nu\displaystyle{\not{\!\!A}}\,\psi^\nu\nonumber\\&&-(\partial\cdot
\bar\psi)\displaystyle{\not{\!\!A}}\,\partial\cdot\psi+2(\partial\cdot A)\left(\bar\psi^\mu\partial_\mu\displaystyle{
\not{\!\psi}}+\displaystyle{\not{\!\bar\psi}}\,\overset\leftarrow{\partial}\,^\mu\psi_\mu\right)+d(...).\eea{com16}
Notice that the 2nd, 4th and 5th terms combine into a total derivative. One can rewrite the 1st and 3rd terms in terms
of the photon field strength by using $\partial_\mu A_\nu=F_{\mu\nu}+\partial_\nu A_\mu$. Also, one can extract a
$\Delta$-exact piece, by using EoMs: $\displaystyle{\not{\!\partial}}\,\psi_\mu-\partial_\mu\displaystyle{\not{\!\psi}}=0$,
in the term containing $(\partial\cdot A)$. This leaves us with \bea V^{(2)}_{\text{ST}}&\approx&2\left(\bar\psi_\alpha
\gamma^\mu\partial^\alpha\psi^\nu-\bar\psi^\mu\overset\leftarrow{\partial}\,^\alpha\gamma^\nu\psi_\alpha\right)F_{\mu\nu}
-2\left[\left(\bar\psi_\alpha\gamma^\mu\partial^\alpha\psi^\nu\right)\partial_\mu A_\nu+\text{h.c.}\right]\nonumber\\
&&+2(\partial\cdot A)\left(\bar\psi^\mu\displaystyle{\not{\!\partial}}\,\psi_\mu+\bar\psi^\mu\overset\leftarrow{\displaystyle
\not{\!\partial}}\psi_\mu\right).\eea{com17} Again, we integrate by parts the last term of the first line, so that no
derivatives act on the photon field. In the second line as well we perform integration by parts to have 2 derivatives
acting on the photon field. This finally gives \bea V^{(2)}_{\text{ST}}&\approx&2\left(\bar\psi_\alpha\gamma^\mu\partial^
\alpha\psi^\nu-\bar\psi^\mu\overset\leftarrow{\partial}\,^\alpha\gamma^\nu\psi_\alpha\right)F_{\mu\nu}\nonumber\\&&+2\left[
\left(\bar\psi_\alpha\partial^\alpha\displaystyle{\not{\!\partial}}\,\psi^\nu+\bar\psi_\alpha\overset\leftarrow{\displaystyle
{\not{\!\partial}}}\,\partial^\alpha\psi^\nu\right)A_\nu+\text{h.c.}\right]-2\bar\psi_\alpha\left(\displaystyle{\not{
\!\partial}}\,\partial\cdot A\right)\psi^\alpha.\eea{com18} It is clear that, in the TT gauge, both the off-shell
vertices~(\ref{com14}) and~(\ref{com18}) reduce to \beq V^{(2)}_{\text{TT}}=2\left(\bar\psi_\alpha\gamma^\mu\partial^
\alpha\psi^\nu-\bar\psi^\mu\overset\leftarrow{\partial}\,^\alpha\gamma^\nu\psi_\alpha\right)F_{\mu\nu},\eeq{com19}
which is nothing but the 2-derivative on-shell vertex given in~\cite{Metsaev}. The equivalence of the two off-shell
vertices is also evident as, upon subtracting~(\ref{com18}) from~(\ref{com14}), we have \beq V^{(2)}-V^{(2)}_{\text{ST}}
=2\bar\psi_\alpha\gamma^\mu\left(\partial^\nu F_{\mu\nu}\right)\psi^\alpha=\Delta\text{-exact}.\eeq{com20}

Finally, we consider the vertex with 3 derivatives, which reads \beq V^{(3)}=\bar{\Psi}_{\mu\alpha}\Psi^\alpha_{~\nu}
F^{\mu\nu}=\left(\partial_\mu\bar\psi^\alpha-\partial^\alpha\bar\psi_\mu\right)\left(\partial_\alpha\psi_\nu-\partial_\nu
\psi_\alpha\right)F^{\mu\nu}.\eeq{com21} Integration by parts w.r.t. $\partial_\mu$, appearing in the 1st term inside the
first parentheses, gives \beq V^{(3)}\approx-\bar\psi^\alpha\left(\partial_\mu\partial_\alpha\psi_\nu-\partial_\mu\partial
_\nu\psi_\alpha\right)F^{\mu\nu}-\bar\psi^\alpha\Psi_{\alpha\nu}\partial_\mu F^{\mu\nu}-\bar\psi_\mu\overset\leftarrow
{\partial}\,^\alpha\partial_\alpha\psi_\nu F^{\mu\nu}+\bar\psi_\mu\overset\leftarrow{\partial}\,^\alpha\partial_\nu\psi_
\alpha F^{\mu\nu}.\eeq{com22} Here the 2nd term inside the parentheses on the right side is identically zero, while the term
containing $\partial_\mu F^{\mu\nu}$ is $\Delta$-exact. We use the 3-box rule~(\ref{3box}) in the penultimate term. Also
we integrate by parts w.r.t. $\partial_\nu$ in the last term, and it produces a $\Delta$-exact piece, containing $\partial_\nu
F^{\mu\nu}$, that we discard. The result is
\bea V^{(3)}\approx-\left(\bar\psi^\alpha\partial_\alpha\partial_\mu\psi_\nu+\bar\psi_\mu\overset\leftarrow{\partial}_\nu
\overset\leftarrow{\partial}_\alpha\psi^\alpha\right)F^{\mu\nu}+\tfrac{1}{2}\left[\bar\psi_\mu\Box\psi_\nu+\bar\psi_\mu
\overset\leftarrow\Box\psi_\nu-\Box\left(\bar\psi_\mu\psi_\nu\right)\right]F^{\mu\nu}.\eea{com23} Now, one can perform a
double integration by parts in the last term in the brackets in order to have $\Box F^{\mu\nu}$, which gives a
$\Delta$-exact piece, so that we finally have
\bea V^{(3)}\approx-\left(\bar\psi^\alpha\partial_\alpha\partial_\mu\psi_\nu+\bar\psi_\mu\overset\leftarrow{\partial}_\nu
\overset\leftarrow{\partial}_\alpha\psi^\alpha\right)F^{\mu\nu}+\tfrac{1}{2}\left(\bar\psi_\mu\Box\psi_\nu+\bar\psi_\mu
\overset\leftarrow\Box\psi_\nu\right)F^{\mu\nu}.\eea{com24} On the other hand, the 3-derivative off-shell ST vertex contains
as many as 14 terms: \bea V^{(3)}_{\text{ST}}=(\partial_\alpha\bar\psi^\mu)(\partial_\mu\psi^\nu)\partial_\nu A^\alpha
-\bar\psi^\mu(\partial_\alpha\partial_\mu\psi^\nu)\partial_\nu A^\alpha-(\partial_\alpha\partial_\nu\bar\psi^\mu)
(\partial_\mu\psi^\nu)A^\alpha+(\partial_\nu\bar\psi^\mu)(\partial_\mu\partial_\alpha\psi^\nu)A^\alpha\nonumber\\
-(\partial_\alpha\bar\psi^\mu)\psi^\nu\partial_\mu\partial_\nu A^\alpha+\bar\psi^\mu(\partial_\alpha\psi^\nu)\partial_\mu
\partial_\nu A^\alpha+(\partial_\alpha\partial_\nu\bar\psi^\mu)\psi^\nu\partial_\mu A^\alpha-(\partial_\nu\bar\psi^\mu)
(\partial_\alpha\psi^\nu)\partial_\mu A^\alpha~~\nonumber\\+(\partial_\mu\partial\cdot\bar\psi)(\partial\cdot\psi)A^\mu
-(\partial\cdot\bar\psi)(\partial_\mu\partial\cdot\psi)A^\mu+\bar\psi^\mu(\partial_\mu\partial_\alpha\psi^\alpha)\partial
\cdot A~~~~~~~~~~~~~~~~~~~~~~~~~~~\;\nonumber\\-\bar\psi^\mu(\partial\cdot\psi)\partial_\mu\partial_\alpha A^\alpha-
(\partial_\mu\partial_\alpha\bar\psi^\alpha)\psi^\mu\partial\cdot A+(\partial\cdot\bar\psi)\psi^\mu\partial_\mu\partial_
\alpha A^\alpha.~~~~~~~~~~~~~~~~~~~~~~~~~~~~~\;\nonumber\eea{xxxxxx} Here we will perform a number of integrations by
parts. In the first line, we integrate by parts the 1st term w.r.t. $\partial_\alpha$, the 3rd w.r.t. $\partial_\mu$, and
the 4th w.r.t. $\partial_\nu$. In the second line, the 1st, 2nd and 4th terms are integrated by parts respectively w.r.t.
$\partial_\nu$, $\partial_\mu$ and $\partial_\alpha$. In the third line, this is done only on the 3rd term w.r.t.
$\partial_\alpha$. Finally, in the fourth line, the 1st and 3rd terms are integrated by parts w.r.t. both
$\partial_\mu$ and $\partial_\alpha$, while the 2nd one only w.r.t. $\partial_\alpha$. Dropping total derivatives, the
result is
\bea V^{(3)}_{\text{ST}}&\approx&-4\bar\psi^\alpha(\partial_\alpha\partial_\mu\psi_\nu)\partial^\nu A^\mu+4\bar\psi_\mu
\overset\leftarrow{\partial}_\nu\overset\leftarrow{\partial}_\alpha\psi^\alpha\partial^\mu A^\nu+2\left(\bar\psi^\mu
\overset\leftarrow{\partial}_\alpha\psi^\alpha-\bar\psi^\alpha\partial_\alpha\psi^\mu\right)\partial_\mu\partial
\cdot A\nonumber\\&&+2\left(\bar\psi\cdot\overset\leftarrow{\partial}\overset\leftarrow{\partial}_\mu\overset
\leftarrow{\partial}_\alpha\psi^\alpha-\bar\psi^\alpha\partial_\alpha\partial_\mu\partial\cdot\psi\right)A^\mu
+\left(\bar\psi\cdot\overset\leftarrow{\partial}\overset\leftarrow{\partial}_\mu\partial_\alpha\psi^\mu-\bar\psi^\mu
\overset\leftarrow{\partial}_\alpha\partial_\mu\partial\cdot\psi\right)A^\alpha\nonumber\\&&+\left[(\partial_\alpha
\partial\cdot\bar\psi)\partial\cdot\psi+\text{h.c.}\right]A^\alpha+\left[(\partial_\alpha\bar\psi_\mu)\partial\cdot\psi
+\text{h.c.}\right]\partial^\mu A^\alpha.\eea{com25} Let us rewrite the first two terms in the first line
in terms of the photon field strength by using $\partial_\mu A_\nu=F_{\mu\nu}+\partial_\nu A_\mu$, and use the 3-box
rule~(\ref{3box}) in the additional terms. Also, we notice that the last line in~(\ref{com25}) reduces exactly to the
2nd term on the second line, up to a total derivative. Then, the vertex reads
\bea V^{(3)}_{\text{ST}}&\approx&4\left(\bar\psi^\alpha\partial_\alpha\partial_\mu\psi_\nu+\bar\psi_\mu\overset\leftarrow
{\partial}_\nu\overset\leftarrow{\partial}_\alpha\psi^\alpha\right)F^{\mu\nu}+2\left(\bar\psi\cdot\overset\leftarrow
{\partial}\overset\leftarrow{\partial}_\mu\partial_\alpha\psi^\mu-\bar\psi^\mu\overset\leftarrow{\partial}_\alpha
\partial_\mu\partial\cdot\psi\right)A^\alpha\nonumber\\&&+2\left(\bar\psi_\mu\overset\leftarrow{\partial}_\nu\Box\psi^\nu
-\bar\psi^\nu\overset\leftarrow{\Box}\partial_\nu\psi_\mu\right)A^\mu-2\left(
\bar\psi^\mu\overset\leftarrow{\partial}_\alpha\psi^\alpha-\bar\psi^\alpha\partial_\alpha\psi^\mu\right)\left(\Box A_\mu
-\partial_\mu\partial\cdot A\right)\nonumber\\&&+2\left[\bar\psi^\alpha\partial_\alpha\left(\Box\psi_\mu-
\partial_\mu\partial\cdot\psi\right)-\left(\bar\psi_\mu\overset\leftarrow{\Box}-\bar\psi\cdot\overset\leftarrow{\partial}
\overset\leftarrow{\partial}_\mu\right)\overset\leftarrow{\partial}_\alpha\psi^\alpha\right]A^\mu.\eea{com26}
Clearly, the 2nd term in the second line and the entire third line are $\Delta$-exact, while, modulo $\Delta$-exact pieces,
the 1st term in the second line can have $\Box\psi^\nu$ replaced by $\partial^\nu\partial\cdot\psi$. The latter result can
be combined with the 2nd term in the first line to give
\bea V^{(3)}_{\text{ST}}&\approx&4\left(\bar\psi^\alpha\partial_\alpha\partial_\mu\psi_\nu+\bar\psi_\mu\overset\leftarrow
{\partial}_\nu\overset\leftarrow{\partial}_\alpha\psi^\alpha\right)F^{\mu\nu}-2\left(\bar\psi\cdot\overset\leftarrow
{\partial}\overset\leftarrow{\partial}_\mu\Psi^{\mu\nu}-\bar\Psi^{\mu\nu}\partial_\mu\partial\cdot\psi\right)A_\nu\nonumber\\
&\approx&4\left(\bar\psi^\alpha\partial_\alpha\partial_\mu\psi_\nu+\bar\psi_\mu\overset\leftarrow{\partial}_\nu\overset\leftarrow
{\partial}_\alpha\psi^\alpha\right)F^{\mu\nu}+\left(\bar\psi\cdot\overset\leftarrow{\partial}\Psi_{\mu\nu}-\bar\Psi_{\mu\nu}
\partial\cdot\psi\right)F^{\mu\nu},\eea{com27} where we have reached the second step by performing integration by parts
w.r.t. $\partial_\mu$ in the 2nd term of the first step, and dropping $\Delta$-exact terms containing $\partial_\mu\Psi^{\mu\nu}$.
In the 2nd term of the second step, one can write $\Psi_{\mu\nu}=2\partial_{[\mu}\psi_{\nu]}$, and integrate by parts to obtain,
among others, $\Delta$-exact terms containing $\partial_\mu F^{\mu\nu}$, which can be dropped. The result is
\beq V^{(3)}_{\text{ST}}\approx-\left(\bar\psi^\alpha\partial_\alpha\partial_\mu\psi_\nu+\bar\psi_\mu\overset\leftarrow
{\partial}_\nu\overset\leftarrow{\partial}_\alpha\psi^\alpha\right)F^{\mu\nu}+\tfrac{1}{2}\left(\bar\psi_\mu\partial_\nu\partial
\cdot\psi+\bar\psi\cdot\overset\leftarrow{\partial}\overset\leftarrow{\partial}_\mu\psi_\nu\right)F^{\mu\nu},\eeq{com28}
where we have made the rescaling $A_\mu\rightarrow-\tfrac{1}{4}A_\mu$, for convenience of comparison with our vertex $V^{(3)}$.
One finds that both the vertices reduce in the TT gauge to \beq V_{\text{TT}}^{(3)}=-\left(\bar\psi^\alpha\partial_\alpha
\partial_\mu\psi_\nu+\bar\psi_\mu\overset\leftarrow{\partial}_\nu\overset\leftarrow{\partial}_\alpha\psi^\alpha\right)F^{\mu\nu},
\eeq{com29} which indeed is the 3-derivative on-shell vertex reported in~\cite{Metsaev}. In view of Eqn.~(\ref{com24})
and~(\ref{com28}), one also finds that the two vertices differ by $\Delta$-exact terms:
\beq V^{(3)}-V_{\text{ST}}^{(3)}\approx\tfrac{1}{2}\left[\bar\psi_\mu\left(\Box\psi_\nu-\partial_\nu\partial\cdot\psi\right)
+\left(\bar\psi_\mu\overset\leftarrow{\Box}-\bar\psi\cdot\overset\leftarrow{\partial}\overset\leftarrow{\partial}_\mu\right)
\psi_\nu\right]F^{\mu\nu}=\Delta\text{-exact}.\eeq{com30} This shows the equivalence of the full off-shell vertices.

\subsection{1---$\textbf{\textit s}$---$\textbf{\textit s}$ Vertices: $\textbf{\textit s}\geq\textbf{5/2}$}\label{sec:coms}

For the sake of simplicity, from now on we restrict our attention to on-shell equivalence of vertices. As we already mentioned,
if two vertices match in some gauge, say the TT one, they should also be off-shell equivalent. With this end in view, we read
off the ST vertices~\cite{Taronna}, which would generally contain a bunch of terms to begin with even in the TT gauge. However,
one can perform integrations by parts to see that actually the on-shell vertices are extremely simple, containing no more than
a few non-trivial terms.

For example, one can take the 3-derivative $1-\tfrac{5}{2}-\tfrac{5}{2}$ ST vertex in the TT gauge, and integrate by parts in
order to have one derivative on each field. The result is simply
\beq
V^{(3)}_{\text{ST}}\sim\bar{\psi}_{\mu\alpha}\overset\leftarrow\partial_\beta F^{\mu\nu}\partial^\alpha\psi^\beta_{~\nu}
+\bar{\psi}_{\mu\alpha}\overset\leftarrow\partial_\rho\left(\partial_\beta A^\rho\right)\partial^\alpha\psi^{\beta\mu},
\eeq{3derspin5/2}
where $\sim$ means equivalence in the TT gauge up to an overall factor. In the 2nd term we integrate by parts to avoid
derivatives on the photon field. We get
\beq
V^{(3)}_{\text{ST}}\sim\bar{\psi}_{\mu\alpha}\overset\leftarrow\partial_\beta F^{\mu\nu}\partial^\alpha\psi^\beta_{~\nu}-\bar
{\psi}_{\mu\alpha}\overset\leftarrow\partial_\beta\left(\overset\leftarrow\partial\cdot A\right)\partial^\alpha\psi^{\beta\mu}.
\eeq{3derspin5/2-1}
One can make use of the Clifford algebra to write $\overset\leftarrow\partial\cdot A=\tfrac{1}{2}\overset\leftarrow\partial_\rho
A_\sigma\left(\gamma^\rho\gamma^\sigma+\gamma^\sigma\gamma^\rho\right)$, in the 2nd term on the right hand side of
Eqn.~(\ref{3derspin5/2-1}), and then integrate by parts w.r.t. this derivative. Dropping some $\Delta$-exact terms in the TT gauge,
we get
\beq
V^{(3)}_{\text{ST}}\sim\bar{\psi}_{\mu\alpha}\overset\leftarrow\partial_\beta F^{\mu\nu}\partial^\alpha\psi^\beta_{~\nu}
+\tfrac{1}{2}\bar{\psi}_{\mu\alpha}\overset\leftarrow\partial_\beta\left(\partial_\rho A_\sigma\right)\gamma^\sigma
\gamma^\rho\partial^\alpha\psi^{\beta\mu}.
\eeq{3derspin5/2-2}
Because $\partial\cdot A=0$ in our gauge choice, we can write $\left(\partial_\rho A_\sigma\right)\gamma^\sigma
\gamma^\rho=-\tfrac{1}{2}\displaystyle{\not{\!F}}$, by making use of the identity $\gamma^\sigma\gamma^\rho=
\eta^{\sigma\rho}-\gamma^{\sigma\rho}$. Therefore, we are left with
\beq
V^{(3)}_{\text{ST}}\sim\bar{\psi}_{\mu\alpha}\overset\leftarrow\partial_\beta\left(F^{\mu\nu}-\tfrac{1}{4}\eta^{\mu\nu}
\displaystyle{\not{\!F}}\right)\partial^\alpha\psi^\beta_{~\nu}\,.
\eeq{3derspin5/2-3}
We would like to see how this compares with our 3-derivative $1-\tfrac{5}{2}-\tfrac{5}{2}$ vertex,
\beq
V^{(3)}=\bar{\psi}^{\,(1)}_{\alpha\beta\|\,\mu}\,F^{+\mu\nu}\psi^{(1)\alpha\beta\|}_{~~~~~~~\nu}\,.
\eeq{3derspin5/2our}
The same steps as took us from Eqn.~(\ref{com1}) to Eqn.~(\ref{com2}) for the spin-$\tfrac{3}{2}$ case lead to
\beq
V^{(3)}\sim2\,\bar{\psi}^{\,(1)}_{\alpha\beta\|\,\mu}\left(F^{\mu\nu}-\tfrac{1}{4}\eta^{\mu\nu}
\displaystyle{\not{\!F}}\right)\psi^{(1)\alpha\beta\|}_{~~~~~~~\nu}\,.
\eeq{3derspin5/2our-1}
Now, one can rewrite the fermionic $1$-curl in terms of the original field. There will be terms that have at least
one pair of mutually contracted derivatives: one acting on $\bar{\psi}_\mu$ and the other on $\psi_\mu$. For such
terms  one can make use of the 3-box rule~(\ref{3box}) to see that they are trivial in the TT gauge. Up to a trivial
factor, one then has
\beq
V^{(3)}\sim\bar{\psi}_{\mu\alpha}\overset\leftarrow\partial_\beta\left(F^{\mu\nu}-\tfrac{1}{4}\eta^{\mu\nu}
\displaystyle{\not{\!F}}\right)\partial^\alpha\psi^\beta_{~\nu}\,.
\eeq{3derspin5/2our-2}
From Eqn.~(\ref{3derspin5/2-3}) and~(\ref{3derspin5/2our-2}), we see that the two vertices are indeed on-shell
equivalent.

Let us move on to the $4$-derivative $1-\tfrac{5}{2}-\tfrac{5}{2}$ vertex. The ST one is found to be
\beq
V^{(4)}_{\text{ST}}\sim\bar{\psi}_{\mu\nu}\overset\leftarrow{\partial}_\rho\overset\leftarrow{\partial}_\sigma
\displaystyle{\not{\!\!A}}\,\partial^\mu\partial^\nu\psi^{\rho\sigma},
\eeq{4derST}
whereas our one is given by
\beq
V^{(4)}=\left(\bar{\Psi}_{\mu\nu|\rho\sigma}\,\gamma^{\mu\nu\alpha\beta\lambda}\,\Psi_{\alpha\beta|}^{~~~\,\rho\sigma}
\right)A_\lambda\approx-2\left(\bar{\Psi}_{\mu\nu|\rho\sigma}\,\gamma^\lambda\,\Psi^{\mu\nu|\rho\sigma}\right)A_\lambda.
\eeq{4derours}
We rewrite the curvature in terms of the spin-$\tfrac{5}{2}$ field. Among the resulting terms those with contracted
pair(s) of derivatives are, again, trivial in the TT gauge, thanks to the 3-box rule~(\ref{3box}). The other terms
clearly add up to reproduce the expression~(\ref{4derST}). Therefore,
\beq
V^{(4)}\approx V^{(4)}_{\text{ST}}\sim\bar{\psi}_{\mu\nu}\overset\leftarrow{\partial}_\rho\overset\leftarrow
{\partial}_\sigma\displaystyle{\not{\!\!A}}\,\partial^\mu\partial^\nu\psi^{\rho\sigma}.
\eeq{4dercompare}

For spin $\tfrac{5}{2}$, the only other vertex is the $5$-derivative one. The ST one reads
\beq
V^{(5)}_{\text{ST}}\sim\left(\bar{\psi}_{\mu\nu}\overset\leftarrow{\partial}_\rho\overset\leftarrow{\partial}
_\sigma\overset\leftrightarrow{\partial}\,^\lambda\,\partial^\mu\partial^\nu\psi^{\rho\sigma}\right)A_\lambda.
\eeq{5derST}
On the other hand, we have the 5-derivative Born-Infeld type vertex:
\beq
V^{(5)}=\bar{\Psi}_{\alpha\beta|\,\mu\rho}\Psi^{\alpha\beta|\,\rho}_{~~~~~\nu}F^{\mu\nu}\approx\tfrac{1}{2}
\left(\bar{\Psi}_{\mu\nu|\,\rho\sigma}\,\overset\leftrightarrow{\partial}\,^\lambda\,\Psi^{\mu\nu|\,\rho\sigma}\right)
A_\lambda.\eeq{5derours} The off-shell equivalence can be understood in view of Eqn.~(\ref{rs23})--(\ref{rs25}),
which pertain to spin $\tfrac{3}{2}$. In the equivalent vertex, again, we rewrite the fermionic curvature in terms
of the spin-$\tfrac{5}{2}$ field, and massage the resulting terms the same way as was done for $V^{(4)}$.
Thus, up to overall factors, we reproduce on-shell~(\ref{5derST}), so that
\beq
V^{(5)}\approx V^{(5)}_{\text{ST}}\sim\left(\bar{\psi}_{\mu\nu}\overset\leftarrow{\partial}_\rho\overset\leftarrow
{\partial}_\sigma\overset\leftrightarrow{\partial}\,^\lambda\,\partial^\mu\partial^\nu\psi^{\rho\sigma}\right)A_\lambda.
\eeq{5dercompare}

For arbitrary spin, $s=n+\tfrac{1}{2}$, the story is very similar, and there are no further complications other than
cluttering of indices. One can write down the ST vertices in the TT gauge from Eqn.~(A.16) of~\cite{Taronna}. They turn
out to be
\bea
V^{(2n-1)}_\textrm{ST}&\sim&\bar{\psi}_{\mu\,\alpha_1\dots\alpha_{n-1}}\overset\leftarrow{\partial}_{\beta_1}...\overset
\leftarrow{\partial}_{\beta_{n-1}}F^{\mu\nu}\partial^{\alpha_1}...\partial^{\alpha_{n-1}}\psi^{\beta_1\dots\beta_{n-1}}
_{~~~~~~~~~\;\nu}\nonumber\\&&-\bar{\psi}_{\mu\,\alpha_1\dots\alpha_{n-1}}\overset\leftarrow{\partial}_{\beta_1}...
\overset\leftarrow{\partial}_{\beta_{n-1}}\left(\overset\leftarrow{\partial}\cdot A\right)\partial^{\alpha_1}...
\partial^{\alpha_{n-1}}\psi^{\beta_1\dots\beta_{n-1}\,\mu},\label{arb2n-1}\\V^{(2n)}_\textrm{ST}&\sim&\bar{\psi}_{\mu_1
\dots\mu_n}\overset\leftarrow{\partial}_{\nu_1}...\overset\leftarrow{\partial}_{\nu_n}\displaystyle{\not{\!\!A}}\,
\partial^{\mu_1}...\partial^{\mu_n}\psi^{\nu_1\dots\nu_n},\label{arb2n}\\ V^{(2n+1)}_\textrm{ST}&\sim&
\left(\bar{\psi}_{\mu_1\dots\mu_n}\overset\leftarrow{\partial}_{\nu_1}...\overset\leftarrow{\partial}_{\nu_n}\overset
\leftrightarrow{\partial}\,^\lambda\,\partial^{\mu_1}...\partial^{\mu_n}\psi^{\nu_1\dots\nu_n}\right)A_\lambda.
\eea{arb2n+1} Their similarity with the lower-spin counterparts is obvious. Indeed, setting $n=2$ produces exactly the
respective $1-\tfrac{5}{2}-\tfrac{5}{2}$ vertices given in Eqn.~(\ref{3derspin5/2-1}),~(\ref{4derST}) and~(\ref{5derST}).
One can massage the $(2n-1)$-derivative vertex, in particular, the same way as its spin-$\tfrac{5}{2}$ counterpart to
obtain an arbitrary-spin generalization of Eqn.~(\ref{3derspin5/2-3}), namely
\beq
V^{(2n-1)}_{\text{ST}}\sim\bar{\psi}_{\mu\,\alpha_1\dots\alpha_{n-1}}\overset\leftarrow{\partial}_{\beta_1}...
\overset\leftarrow{\partial}_{\beta_{n-1}}\left(F^{\mu\nu}-\tfrac{1}{4}\eta^{\mu\nu}\displaystyle{\not{\!F}}\right)
\partial^{\alpha_1}...\partial^{\alpha_{n-1}}\psi^{\beta_1\dots\beta_{n-1}}_{~~~~~~~~~\;\nu}\,.
\eeq{arb2n-1-1}
Our arbitrary-spin vertices are also straightforward generalizations of their lower-spin examples. In view of
the spin-$\tfrac{5}{2}$ counterparts, Eqn.~(\ref{3derspin5/2our}),~(\ref{4derours}) and~(\ref{5derours}), one can write
\bea
V^{(2n-1)}&\approx&\bar{\psi}^{(n-1)}_{\alpha_1\beta_1|...|\alpha_{n-1}\beta_{n-1}\|\,\mu}\,F^{+\mu\nu}\psi^{(n-1)
\alpha_1\beta_1|...|\alpha_{n-1}\beta_{n-1}\|}_{~~~~~~~~~~~~~~~~~~~~~~~~~\,\nu}\,,\label{arbour2n-1}\\
V^{(2n)}&\approx&\left(\bar{\Psi}_{\mu_1\nu_1|\dots|\mu_n\nu_n}\gamma^\lambda\,\Psi^{\mu_1\nu_1|\dots|\mu_n\nu_n}
\right)A_\lambda,\label{arbour2n}\\
V^{(2n+1)}&\approx&\left(\bar{\Psi}_{\mu_1\nu_1|\dots|\mu_n\nu_n}\overset\leftrightarrow{\partial}\,^\lambda\,
\Psi^{\mu_1\nu_1|\dots|\mu_n\nu_n}\right)A_\lambda.
\eea{arbour2n+1}
Again, one can use the 2nd identity in~(\ref{id4}) to rewrite the $F^{+\mu\nu}$ in the first vertex, and express the
fermionic ($n-1$)- and $n$-curls in all the vertices~(\ref{arbour2n-1})--(\ref{arbour2n+1}) in terms of the original
field. The terms with contracted pair(s) of derivatives are, as usual, subject to the 3-box rule~(\ref{3box}), and
hence trivial in the TT gauge. One finds that our vertices indeed reduce on shell respectively to~(\ref{arb2n-1-1}),
~(\ref{arb2n}) and~(\ref{arb2n+1}). This proves the on-shell (and therefore off-shell) equivalence of the $1-s-s$
vertices: \beq V^{(p)}\sim V^{(p)}_{\text{ST}},\qquad p=2n-1,\,2n,\,2n+1.\eeq{arbcompare}

\section{Second-Order Consistency}\label{sec:2nd}

We recall that consistent second-order deformation requires $(S_1,S_1)$ to be $s$-exact: \beq (S_1,S_1)=-2sS_2=
-2\Delta S_2 -2\Gamma S_2.\eeq{2nd1} For abelian vertices, this antibracket is zero, so that the first-order deformations
always go unobstructed. Non-abelian vertices, however, are more interesting in this respect.

We can see that there is obstruction for the non-abelian vertices we have obtained, which do \emph{not} obey Eqn.~(\ref{2nd1}).
We prove our claim by contradiction. If Eqn.~(\ref{2nd1}) holds, then the most general form of the antibracket evaluated
at zero antifields is \beq \left[(S_1,S_1)\right]_{\Phi^*_A=0} = \Delta M + \Gamma N,\eeq{2nd2} where $M=-2\left[S_2\right]
_{\mathcal{C}^{*}_{\alpha}=0}$ and $N=-2\left[S_2\right]_{\Phi^*_A=0}$. Note that $M$ is obtained by setting to zero
\emph{only} the antighosts in $S_2$. Furthermore, the equality~(\ref{2nd2}) holds precisely because $S_2$ is linear in the
antifields. The $\Gamma$-variation of~(\ref{2nd2}) is therefore $\Delta$-exact:
\beq \Gamma\left[(S_1,S_1)\right]_{\Phi^*_A=0}=\Gamma\Delta M=-\Delta\left(\Gamma M\right).\eeq{2nd3}

It is relatively easier to compute the left hand side of~(\ref{2nd3}) for our non-abelian vertices. For spin $\tfrac{3}{2}$, we recall
that \bea &a_2=-C^*\bar{\xi}\xi,\qquad a_1=A^{*\mu}(\bar{\psi}_\mu\xi-\bar{\xi}\psi_\mu)+\tilde{a}_1,\qquad a_0=\bar{\psi}_\mu F^{+\mu\nu}\psi_\nu,&\label{2nd4}\\&\tilde{a_1}=i\left[\bar{\psi}^{*\mu}\gamma^\nu F_{\mu\nu}-\tfrac{1}{2(D-2)}\displaystyle
{\not{\!\bar{\psi}^*}}\displaystyle{\not{\!F}}\right]\xi+\text{h.c.}\eea{2nd5} To compute the antibracket of $S_1=\int(a_2+a_1+a_0)$
with itself, we notice that a field-antifield pair shows up only in $\int a_1$, and between $\int a_0$ and $\int a_1$, so that
it reduces to \beq \left(S_1,S_1\right)=2\left(\int a_0,\int a_1\right)+\left(\int a_1,\int a_1\right).\eeq{2nd6} Now, the
second antibracket on the right hand side necessarily contains antifields, while the first one will not contain any. Thus we
have \beq \left[\left(S_1,S_1\right)\right]_{\Phi^*_A=0}=2\left(\int a_0,\int a_1\right).\eeq{2nd7} Notice that, while the
unambiguous piece in $a_1$ contains the antifield $A^{*\mu}$, the ambiguity, $\tilde a_1$, contains instead the antifield
$\bar{\psi}^{*\mu}$. Correspondingly, $\left(\int a_0,\int a_1\right)$  will contain two distinct kinds of pieces: 4-Fermion
terms and Fermion bilinears. Explicitly, \bea \left[\left(S_1,S_1\right)\right]_{\Phi^*_A=0}&=&\int d^Dx\left\{4(\bar{\psi}_\mu
\xi-\bar{\xi}\psi_\mu)\,\partial_\nu\left(\bar{\psi}^{[\mu}\psi^{\nu]}+\tfrac{1}{2}\bar{\psi}_\alpha\gamma^{\mu\nu\alpha\beta}
\psi_\beta\right)\right\}\nonumber\\&&+\int d^Dx\left\{i\bar{\psi}_\mu F^{+\mu\nu}\left[2\gamma^\rho F_{\nu\rho}-\tfrac{1}
{(D-2)}\gamma_\nu\displaystyle{\not{\!F}}\right]\xi+\text{h.c.}\right\}.\eea{2nd8} If the vertex is unobstructed and
Eqn.~(\ref{2nd3}) holds, then the $\Gamma$-variation of each of these terms should independently be $\Delta$-exact. Let us
consider the Fermion bilinears appearing in the second line of Eqn.~(\ref{2nd8}), originating from
$\left(\int a_0,\int\tilde a_1\right)$. It is easy to see that their $\Gamma$-variation is \emph{not} $\Delta$-exact.
We conclude that the non-abelian $1-\tfrac{3}{2}-\tfrac{3}{2}$ vertex gets obstructed beyond the cubic order. The proof
for arbitrary spin will be very similar.

Notice that the vertex $\bar{\psi}_\mu F^{+\mu\nu}\psi_\nu$ is precisely the Pauli term appearing in $\mathcal N=2$
SUGRA~\cite{SUGRA}. The theory, however, contains additional degrees of freedom, namely graviton, on top of a complex massless
spin $\tfrac{3}{2}$ and a $U(1)$ field. It is this new dynamical field that renders the vertex unobstructed, while keeping
locality intact. If one decouples gravity by taking $M_\text{P}\rightarrow\infty$, the Pauli term vanishes because the
dimensionful coupling constant is nothing but $1/M_\text{P}$~\cite{SUGRA}. One could integrate out the \emph{massless} graviton
to obtain a system of spin-$\tfrac{3}{2}$ and spin-1 fields only. The resulting theory contains the Pauli term, but is necessarily
\emph{non-local}. Thus, higher-order consistency of the non-abelian vertex is possible either by forgoing locality or by adding
a new dynamical field (graviton).

\section{Remarks \& Future Perspectives}\label{sec:remarks}

In this paper, we have employed the BRST-BV cohomological methods to construct consistent parity-preserving off-shell
cubic vertices for fermionic gauge fields coupled to EM in flat space. We have shown that consistency and non-triviality
of the deformations forbid minimal coupling, and pose number-of-derivative restrictions on a $1-s-s$ vertex, in
accordance with Metsaev's light-cone-formulation results~\cite{Metsaev}.

The vertices either deform the gauge algebra or, when they do not deform the gauge algebra, turn out not to deform the
gauge transformations either and to deform only the Lagrangian. The non-abelian ones get obstructed in a local theory
beyond the cubic order in the absence of additional higher-spin gauge fields.

Our off-shell cubic vertices are equivalent to the string-theory-inspired ones of Sagnotti-Taronna~\cite{Taronna}.
Note that in~\cite{Taronna} there appears just one dimensionful coupling constant, which can be set to unity.
Then, each cubic vertex will come with a fixed known numerical coefficient. This is apparently in contrast with
our results, where each of the three $1-s-s$ vertices has an independent coupling constant. However, it is well
known that higher-order consistency requirements may impose restrictions on the cubic couplings by relating them
with one another~\cite{BBH}. Because the consistency of string theory is not limited to the cubic order, then it
should not come as a surprise that the cubic vertices it gives rise to have no freedom in the coupling constant.
At any rate, string theory may not be the unique consistent theory of higher-spin fields. If this is true, other
possible choices of the cubic couplings would pertain to other consistent theories.

The number of possible $1-s-s$ vertices for fermions differ from that for bosons. While in both cases there is
only one non-abelian vertex, fermions have, beside the usual Born-Infeld type 3-curvature term, another abelian
vertex in $D\geq5$, which is gauge invariant up to a total derivative. In this respect, fermionic $1-s-s$
vertices are strikingly similar in nature to the bosonic $2-s-s$ ones. The latter include one non-abelian
$(2s-2)$-derivative vertex, a $2s$-derivative abelian one that is gauge invariant up to a total derivative
and exists in $D\geq5$, and a Born-Infeld type abelian one containing $2s+2$ derivatives. This could be seen
by using either the light-cone method~\cite{Metsaev_Boson} or the cohomological methods~\cite{2-3-3,2-s-s}.

For gravitational coupling, spin $\tfrac{3}{2}$ has no consistency issues, but consistent deformations of the
free theory uniquely lead one to $\mathcal N=1$ SUGRA~\cite{N=1SUGRA}, under certain reasonable assumptions.
Fermionic gauge fields with higher spin, $s\geq\tfrac{5}{2}$, and their coupling to gravity are more interesting,
for which one can also employ the BRST-deformation technique~\cite{GHR}. Another interesting avenue to pursue are
the mixed-symmetry fields.

It is instructive to consider the EM coupling of \emph{massive} higher-spin fields in flat space, which has been
discussed by various authors\,\footnote{See, for example, Ref.~\cite{Rakib,Zinoviev,Massive,PR1} and references therein.}.
If Lorentz, parity and time-reversal symmetries hold good, a massive spin-$s$ particle will have $2s+1$ EM
multipoles~\cite{Massive}. This immediately sets for the possible number of derivatives in a $1-s-s$ vertex an upper
bound, which remains the same in the massless limit. The assumption of light-cone helicity conservation in $D=4$
uniquely determines all the multipoles~\cite{Massive}. However, only the highest multipole survives in an appropriate
massless chargeless scaling limit. This observation is in harmony with our results, since any of our lower-derivative
vertices either vanishes in 4D or is not consistent by itself in a local theory.

On the other hand, causal propagation of a charged massive field may call for certain non-minimal terms. Indeed, for a
massive spin $\tfrac{3}{2}$ in flat space, causality analysis in a constant external EM background~\cite{PR1} or in the
case of $\mathcal{N}=2$ \emph{broken} SUGRA~\cite{BrokenSUGRA} reveals the crucial role played by the Pauli term,
$\bar\psi_\mu F^{+\mu\nu}\psi_\nu$. In the massless case, as we have seen, the same term arises as the unique non-abelian
deformation of the gauge theory. These facts go in favor of the gauge-invariant (St\"uckelberg-invariant) formulation,
adopted in~\cite{Zinoviev}, for constructing consistent EM interactions of massive higher spins.

Our non-abelian vertices are seen to be inconsistent beyond the cubic order in a local theory. Such obstructions are rather
common for massless higher-spin vertices in flat space, and some could not even be removed by the inclusion of an
(in)finite number of higher-spin fields, as has been argued in~\cite{Obstruction}. Non-locality may therefore be essential.
In fact, as noticed in~\cite{4-point}, evidence for non-locality shows up already at the quartic level. The
geometric formulation of \emph{free} massless higher spins also hints towards the same, as they generically yield non-local
EoMs~\cite{Dario} if higher-derivative terms are not considered~\cite{BB}.

If one has to give up locality, what becomes relevant for studying higher-spin interactions is a formulation that does not
require locality as an input, e.g., the old S-matrix theory, or perhaps the more powerful BCFW construction~\cite{BCFW} and
generalizations thereof. The latter seem promising for the systematic search of consistent interactions of massless
higher-spin particles in 4D Minkowski space~\cite{Benincasa}.

There are certain technical difficulties in extending the applicability of the BRST-BV cohomological methods to constant
curvature spaces. For AdS space, in particular, those could be avoided by using the ambient-space formulation~\cite{Ambient}.
If so, one would be able to construct off-shell vertices for AdS, and compare them with the recently-obtained results
of~\cite{Cubic_AdS0,Cubic_AdS}. This would be one step towards finding a standard action for the Vasiliev systems~\cite{Vasiliev},
which are a consistent set of non-linear equations for symmetric tensors of arbitrary rank in any number of dimensions.
We leave this as future work.

\subsection*{Acknowledgments}

We would like to thank N.~Boulanger, D.~Francia, E.~Joung, K.~Mkrtchyan, M.~Porrati and M.~Taronna for useful discussions.
MH and RR are grateful to The Erwin Schr\"odinger Institute, Vienna for its kind hospitality, during the Workshop on
Higher Spin Gravity, while this work was ongoing. MH gratefully acknowledges support from the Alexander von Humboldt
Foundation through a Humboldt Research Award and support from the ERC through the ``SyDuGraM" Advanced Grant.
RR is a Postdoctoral Fellow of the Fonds de la Recherche Scientifique-FNRS. This work is also partially supported by
IISN - Belgium (conventions 4.4511.06 and 4.4514.08) and by the ``Communaut\'e Fran\c{c}aise de Belgique" through the
ARC program.

\begin{appendix}
\numberwithin{equation}{section}

\section{Curvatures \& Equations of Motion}\label{sec:Curvature}

Let us recall that for arbitrary spin $s=n+\tfrac{1}{2}$, we have a totally symmetric rank-$n$ tensor-spinor
$\psi_{\mu_1...\mu_n}$, whose curvature is its $n$-curl, i.e., the rank-$2n$ tensor
\beq \Psi_{\mu_1\nu_1|\mu_2\nu_2|...|\mu_n\nu_n}=\left[...\left[\,\left[\partial_{\mu_1}...
\partial_{\mu_n}\psi_{\nu_1...\nu_n}-(\mu_1\leftrightarrow\nu_1)\right]-(\mu_2\leftrightarrow\nu_2)\right]...
\right]-(\mu_n\leftrightarrow\nu_n).\eeq{curv1} Notice that, unlike the Fronsdal tensor, $\mathcal S_{\mu_1...
\mu_n}$, the curvature tensor~(\ref{curv1}) is gauge invariant even for an \emph{unconstrained} gauge parameter.
Its properties can be found in~\cite{Curvature}. The curvature is antisymmetric under the interchange of ``paired''
indices, e.g., \beq \Psi_{\mu_1\nu_1|\mu_2\nu_2|\dots|\mu_n\nu_n}=-\Psi_{\nu_1\mu_1|\mu_2\nu_2|\dots|\mu_n\nu_n},
\eeq{curv2} but symmetric under the interchange of any two sets of paired indices, e.g., \beq \Psi_{\mu_1\nu_1|
\mu_2\nu_2|\dots|\mu_{n-1}\nu_{n-1}|\mu_n\nu_n}=\Psi_{\mu_n\nu_n|\mu_2\nu_2|\dots|\mu_{n-1}\nu_{n-1}|\mu_1\nu_1}.
\eeq{curv3} These symmetries actually hold good for any $m$-curl of the field, $m\leq n$. Another important property
of the curvature is that it obeys the Bianchi identity \beq \partial_{[\rho}\Psi_{\mu_1\nu_1]|\mu_2\nu_2|...|\mu_n
\nu_n}=0.\eeq{curv4} The (Weinberg) curvature~(\ref{curv1}) and the EM field strength $F_{\mu\nu}$ are useful in
casting the EoMs into a variety of forms, which, among others, can help one identify $\Delta$-exact pieces.
First, we write down these various forms for the photon field. Next, we do the same for spin $\tfrac{3}{2}$,
explaining as well how to derive them and writing them explicitly as $\Delta$-variations. Then we move on to spin
$\tfrac{5}{2}$, and finally to arbitrary spin.

\subsection{The Photon}

The original photon EoMs are given by \beq \partial^\mu F_{\mu\nu}=\Box A_\nu-\partial_\nu(\partial\cdot A)=\Delta
A^{*}_{\nu}.\eeq{phot1} One can take its 1-curl to obtain \beq\Box F_{\mu\nu}=2\Delta (\partial_{[\mu} A^{*}_{\nu]}).
\eeq{phot2}

\subsection{Spin \textbf{3/2}}

For spin $\tfrac{3}{2}$, the original EoMs can be obtained directly from the master action~(\ref{rs5})
\sea{noname1}{\gamma^{\mu\alpha\beta}\Psi_{\alpha\beta}&=-2i\Delta\psi^{*\mu},\label{eom3/2}\\\bar{\Psi}_{\alpha
\beta}\gamma^{\alpha\beta\mu}&=2i\Delta\bar{\psi}^{*\mu}.\label{eom3/2bar}} One can take the $\gamma$-trace of
Eqn.~(\ref{eom3/2}), and use $\gamma_\mu\gamma^{\mu\alpha\beta}=(D-2)\gamma^{\alpha\beta}$, to obtain
\beq \gamma^{\mu\nu}\Psi_{\mu\nu}=2\left(\displaystyle{\not{\!\partial}}\displaystyle{\not{\!\psi}}-\partial
\cdot\psi\right)=-2i\Delta\left(\tfrac{1}{D-2}\displaystyle{\not{\!\psi}}^*\right).\eeq{eom3/21} Now, in
Eqn.~(\ref{eom3/2}), one can use the identity $\gamma^{\mu\alpha\beta}=\gamma^\mu\gamma^{\alpha\beta}-2
\eta^{\mu[\alpha}\gamma^{\beta]}$, and then the EoM.~(\ref{eom3/21}), to obtain another very useful form
\beq \gamma^\mu\Psi_{\mu\nu}=\displaystyle{\not{\!\partial}}\,\psi_\nu-\partial_\nu\displaystyle{\not{\!\psi}}
=-i\Delta\left(\psi^{*}_\nu-\tfrac{1}{D-2}\gamma_\nu\displaystyle{\not{\!\psi}}^*\right).\eeq{eom3/22}
One can take a curl of the above equation to get
\beq \displaystyle{\not{\!\partial}}\,\Psi_{\mu\nu}=-2i\Delta\left( \partial_{[\mu}\psi_{\nu]}^*-\tfrac{1}{D-2}
\gamma_{[\nu}\partial_{\mu]}\displaystyle{\not{\!\psi}}^*\right).\eeq{eom3/23} Another useful form can be obtained
by applying the Dirac operator on~(\ref{eom3/22}), and then getting rid of $\displaystyle{\not{\!\partial}}
\displaystyle{\not{\!\psi}}$ in the resulting expression by using~(\ref{eom3/21}). The result is \beq
\partial^\mu\Psi_{\mu\nu}=\Box\psi_\nu-\partial_\nu\left(\partial\cdot\psi\right)=-i\Delta\left[\displaystyle
{\not{\!\partial}}\,\psi^{*}_\nu+\tfrac{1}{D-2}\gamma_{\nu\rho}\partial^\rho\displaystyle{\not{\!\psi}}^*\right].
\eeq{eom3/24} Similarly, one could have started
with~(\ref{eom3/2bar}) to derive the following. \bea &\bar{\Psi}_{\mu\nu}\gamma^{\mu\nu}=2\left(\bar\psi\cdot
\overset\leftarrow{\partial}-\displaystyle\not{\!\bar{\psi}}\overset\leftarrow{\displaystyle\not{\!\partial}}
\right)=2i\Delta\left(\tfrac{1}{D-2}\displaystyle{\not{\!\bar{\psi}^*}}\right),&\label{eom3/2bar1}\\
&\bar\Psi_{\mu\nu}\gamma^\nu=\;\displaystyle\not{\!\bar{\psi}}\overset\leftarrow{\partial}_\mu-\bar\psi_\mu
\overset\leftarrow{\displaystyle\not{\!\partial}}\;=i\Delta\left(\bar{\psi}^{*}_\mu-\tfrac{1}{D-2}\displaystyle
{\not{\!\bar{\psi}^*}}\gamma_\mu\right),&\label{eom3/2bar2}\\&\bar{\Psi}_{\mu\nu}\overset\leftarrow{\displaystyle
{\not{\!\partial}}}=2i\Delta\left(\bar{\psi}^{*}_{[\mu}\overset\leftarrow{\partial}_{\nu]}-\tfrac{1}{D-2}
\displaystyle{\not{\!\bar{\psi}^*}}\,\gamma_{[\mu}\overset\leftarrow\partial_{\nu]}\right),&\label{eom3/2bar3}\\
&\bar\Psi_{\mu\nu}\overset\leftarrow\partial\,^\nu=\left(\bar\psi\cdot\overset\leftarrow\partial\right)
\overset\leftarrow\partial_\mu-\bar{\psi}_\mu\overset\leftarrow\Box=i\Delta\left[\bar\psi^*_\mu\,\overset
\leftarrow{\displaystyle{\not{\!\partial}}}+\tfrac{1}{D-2}\displaystyle{\not{\!\bar\psi}}^*\overset\leftarrow
\partial\,^\rho\gamma_{\rho\mu}\right].&\eea{eom3/2bar4}

\subsection{Spin \textbf{5/2}}

For the spin-$\tfrac{5}{2}$ case, let us recall from Section~\ref{sec:fivehalf} that the original EoMs are given by
\sea{nonmame2}{\mathcal R_{\mu\nu}&=\mathcal{S}_{\mu\nu}-\gamma_{(\mu}\displaystyle\not{\!\mathcal{S}}_{\nu)}-\tfrac
{1}{2}\eta_{\mu\nu}\mathcal{S}'=\Delta \psi^{*}_{\mu\nu},\label{eom5/2}\\\bar{\mathcal R}_{\mu\nu}&=\bar{\mathcal S}_{\mu\nu}-\displaystyle\not{\!\bar{\mathcal S}}_{(\mu}\gamma_{\nu)}-\tfrac{1}{2}\eta_{\mu\nu}\bar{\mathcal S}'
=\Delta \bar{\psi}^{*}_{\mu\nu},\label{eom5/2bar}} which one can easily rewrite in terms of the Fronsdal tensor,
\sea{noname3}{\mathcal S_{\nu_1\nu_2}&=i\left[\displaystyle{\not{\!\partial\,}}\psi_{\nu_1\nu_2}-2\partial_{(\nu_1}
\displaystyle{\not{\!\psi\,}}_{\nu_2)}\right]=\Delta\left[\psi^*_{\nu_1\nu_2}-\tfrac{2}{D}\gamma_{(\nu_1}
\displaystyle{\not{\!\psi^{*}}}_{\nu_2)}-\tfrac{1}{D}\eta_{\nu_1\nu_2}{\psi^{*\prime}}\right],\label{eom5/21}\\
\bar{\mathcal{S}}_{\nu_1\nu_2}&=i\left[\bar\psi_{\nu_1\nu_2}\overset\leftarrow{\displaystyle{\not{\!\partial}}}
-2\displaystyle\not{\!\bar\psi}_{(\nu_1}\overset\leftarrow\partial_{\nu_2)}\right]=\Delta\left[\bar{\psi}^{*}_{\nu_1
\nu_2}-\tfrac{2}{D}\displaystyle{\not{\!\bar\psi}}^*_{(\nu_1}\gamma_{\nu_2)}-\tfrac{1}{D}\eta_{\nu_1\nu_2}{\bar{\psi}
^{*\prime}}\right]\label{eom5/2bar1}.} Now we see that the quantity $\gamma^{\mu_1}\Psi_{\mu_1\nu_1|\,\mu_2\nu_2}$
is given by the $1$-curl of the Fronsdal tensor, so that it is $\Delta$-exact as a result of Eqn.~(\ref{eom5/21}):
\bea \gamma^{\mu_1}\Psi_{\mu_1\nu_1|\,\mu_2\nu_2}&=&\displaystyle{\not{\!\partial\,}}\psi_{\mu_2\nu_2\|\nu_1}^{(1)}
-\partial_{\nu_1}\displaystyle{\not{\!\psi\,}}_{\mu_2\nu_2}^{(1)}~=~-i\mathcal{S}^{(1)}_{\mu_2\nu_2\|\nu_1}
\nonumber\\&=&-i\Delta\left[\psi^{*(1)}_{\mu_2\nu_2\|\nu_1}-\tfrac{1}{D}\gamma_{\nu_1}\displaystyle{\not{\!\psi}}
^{*(1)}_{\mu_2\nu_2}+\tfrac{2}{D}\gamma_{[\mu_2}\partial_{\nu_2]}\displaystyle{\not{\!\psi}}^*_{\nu_1}+\tfrac{2}{D}
\eta_{\nu_1[\mu_2}\partial_{\nu_2]}\psi^{*\prime}\right].~~~~~~~~~~\eea{eom5/23}
Contracting this expression with $\gamma^{\nu_1}$ on the left, we obtain another useful form, \beq \gamma^{\mu_1
\nu_1}\Psi_{\mu_1\nu_1|\,\mu_2\nu_2}=2\left[\displaystyle{\not{\!\partial}}\displaystyle{\not{\!\psi}}^{(1)}_{\mu_2
\nu_2}-\partial^\rho\psi^{(1)}_{\mu_2\nu_2\|\rho}\right]=i\displaystyle{\not{\!\mathcal{S}}}^{(1)}_{\mu_2\nu_2}
=-2i\Delta\left[\tfrac{1}{D}\displaystyle{\not{\!\psi}}^{*(1)}_{\mu_2\nu_2}\right].\eeq{eom5/24}
One finds that taking a curl of~(\ref{eom5/23}) gives yet another form,
\bea \displaystyle{\not{\!\partial}}\,\Psi_{\mu_1\nu_1|\,\mu_2\nu_2}&=&-i\Delta\left[\tfrac{1}{2}\psi^{*(2)}_{\mu_1
\nu_1|\,\mu_2\nu_2}+\tfrac{2}{D}\gamma_{[\mu_1}\partial_{\nu_1]}\displaystyle{\not{\!\psi}}^{*(1)}_{\mu_2\nu_2}+
\tfrac{2}{D}\partial_{[\mu_1}\eta_{\nu_1][\mu_2}\partial_{\nu_2]}\psi^{*\prime}+ (\mu_1\nu_1\leftrightarrow\mu_2
\nu_2)\right]\nonumber\\&=&-i\mathcal{S}^{(2)}_{\mu_1\nu_1|\,\mu_2\nu_2}.\eea{eom5/25}
Given Eqn.~(\ref{eom5/23}) and~(\ref{eom5/24}), one can also write
\bea \partial^{\mu_1}\Psi_{\mu_1\nu_1|\,\mu_2\nu_2}&=&-i\Delta\left[\displaystyle{\not{\!\partial}}\left(\psi^{*(1)}
_{\mu_2\nu_2\|\nu_1}+\tfrac{2}{D}\gamma_{[\mu_2}\partial_{\nu_2]}\displaystyle{\not{\!\psi}}^*_{\nu_1}+\tfrac{2}{D}
\eta_{\nu_1[\mu_2}\partial_{\nu_2]}\psi^{*\prime}\right)+\tfrac{1}{D}\gamma_{\nu_1\rho}\partial^\rho\displaystyle
{\not{\!\psi}}^{*(1)}_{\mu_2\nu_2}\right]\nonumber\\&=&-i\displaystyle{\not{\!\partial}}\,\mathcal{S}^{(1)}_{\mu_2
\nu_2\|\nu_1}+\tfrac{i}{2}\partial_{\nu_1}\displaystyle{\not{\!\mathcal{S}}}^{(1)}_{\mu_2\nu_2}.\eea{eom5/26}
Similarly, Eqn.~(\ref{eom5/2bar1}) gives the various forms of the EoMs for the Dirac conjugate spinor.

\subsection{Arbitrary Spin}

For arbitrary spin $s=n+\tfrac{1}{2}$, we recall from Section~\ref{sec:arbitrary} that the original EoMs read
\sea{noname4}{\mathcal{R}_{\mu_1 \dots \mu_n} &=\mathcal S_{\mu_1...\mu_n}-\tfrac{1}{2}n\,\gamma_{(\mu_1}
\displaystyle\not{\!\mathcal S}_{\mu_2...\mu_n)}-\tfrac{1}{4}n(n-1)\,\eta_{(\mu_1\mu_2}\mathcal S^\prime_{\mu_3
...\mu_n)}=\Delta \psi^{*}_{\mu_1 \dots \mu_n},\label{eomarb}\\\bar{\mathcal{R}}_{\mu_1 \dots \mu_n}&= \mathcal
{\bar S}_{\mu_1...\mu_n}-\tfrac{1}{2}n\displaystyle\not{\!\mathcal{\bar S}}_{(\mu_1...\mu_{n-1}}\gamma_{\mu_n)}
-\tfrac{1}{4}n(n-1)\,\eta_{(\mu_1\mu_2}\mathcal{\bar S}^\prime_{\mu_3...\mu_n)}=\Delta \bar{\psi}^{*}_{\mu_1...
\mu_n}.\label{eomarbbar}} One can reexpress the EoMs in terms of the Fronsdal tensor as follows.
\sea{noname5}{\mathcal{S}_{\nu_1...\nu_n}&=\Delta\left[\psi^*_{\nu_1...\nu_n}-\tfrac{n}{2n+D-4}\,\gamma_{(\nu_1}
\displaystyle{\not{\!\psi}}^*_{\nu_2...\nu_n)}-\tfrac{n(n-1)}{2(n+D-2)}\,\eta_{(\nu_1\nu_2}\psi^{*\prime}
_{\nu_3...\nu_n)}\right],\label{eomarb1}\\\bar{\mathcal{S}}_{\nu_1...\nu_n}&=\Delta\left[\bar{\psi}^*_{\nu_1...
\nu_n}-\tfrac{n}{2n+D-4}\,\displaystyle{\not{\!\bar{\psi}}}^*_{(\nu_1...\nu_{n-1}}\gamma_{\nu_n)}-\tfrac{n(n-1)}
{2(n+D-2)}\,\eta_{(\nu_1\nu_2}\bar{\psi}^{*\prime}_{\nu_3...\nu_n)}\right].\label{eomarbbar1}} From the
definition~(\ref{Fronsdal}) of the Fronsdal tensor, it is easy to see that \beq \gamma^{\mu_1}\Psi_{\mu_1\nu_1|
...|\,\mu_n\nu_n}=-i\mathcal{S}^{(n-1)}_{\mu_2\nu_2|...|\mu_n\nu_n ||\,\nu_1},\eeq{eomarb2} whose contraction
with $\gamma^{\nu_1}$ on the left gives \beq \gamma^{\mu_1\nu_1}\Psi_{\mu_1\nu_1|...|\,\mu_n\nu_n}
=i\displaystyle{\not{\!\mathcal S}}^{(n-1)}_{\mu_2\nu_2|...|\,\mu_n\nu_n}.\eeq{eomarb3} Also, a curl of
Eqn.~(\ref{eomarb2}) yields \beq \displaystyle{\not{\!\partial}}\,\Psi_{\mu_1\nu_1|...|\,\mu_n\nu_n }
=-i\mathcal{S}^{(n)}_{\mu_1\nu_1|...|\,\mu_n\nu_n}.\eeq{eomarb4} Finally, from Eqn.~(\ref{eomarb2})
and~(\ref{eomarb3}) one obtains \beq \partial^{\mu_1}\Psi_{\mu_1\nu_1|...|\,\mu_n\nu_n}=-i\displaystyle{\not
{\!\partial}}\,\mathcal{S}^{(n-1)}_{\mu_2\nu_2|...|\mu_n\nu_n||\,\nu_1}+\tfrac{i}{2}\partial_{\nu_1}\displaystyle
{\not{\!\mathcal{S}}}^{(n-1)}_{\mu_2\nu_2|...|\mu_n\nu_n}.\eeq{eomarb5} In view of Eqn.~(\ref{eomarb1}), it is
now straightforward to write the EoMs~(\ref{eomarb2})--(\ref{eomarb5}) as $\Delta$-exact terms. Similar things
follow from Eqn.~(\ref{eomarbbar1}) for the Dirac conjugate spinor.

\section{The Cohomology of $\Gamma$}\label{sec:Gamma1}

This Appendix is devoted to clarifying and providing proofs of the statements about the cohomology of $\Gamma$
appearing in Section~\ref{sec:Gamma}. We recall that the action of $\Gamma$ is defined by \sea{noname7}{\Gamma
A_\mu&=\partial_\mu C,\label{Gammaaction1}\\\Gamma\psi_{\nu_1...\nu_n}&=n\partial_{(\nu_1}\xi_{\nu_2...\nu_n)}.
\label{Gammaaction2}} Note that the non-trivial elements in the cohomology of $\Gamma$ are nothing but
gauge-invariant objects that themselves are not gauge variations of something else. Here we consider one by one
all such elements enlisted in Section~\ref{sec:Gamma}. In the process, we also prove the statements made towards
the end of  Section~\ref{sec:Gamma} about some $\Gamma$-exact terms.

\subsection{The Curvatures}\label{sec:Curvatures}

The curvatures $\{F_{\mu\nu}, \Psi_{\mu_1\nu_1|...|\,\mu_n\nu_n}\}$ and their derivatives belong to the cohomology
of $\Gamma$. Seeing that the curvatures are $\Gamma$-closed is straightforward. For the photon it follows directly
from the commutativity of partial derivatives as one takes a curl of Eqn.~(\ref{Gammaaction1}),
\beq \Gamma F_{\mu\nu}=\Gamma\left(2\partial_{[\mu}A_{\nu]}\right)=2\partial_{[\mu}\partial_{\nu]}C=0.\eeq{gamma1}
On the other hand, taking a 1-curl of Eqn.~(\ref{Gammaaction2}) one obtains \beq \Gamma\,\psi^{(1)\mu_1\nu_1\|}
_{~~~~~~~~\,\nu_2...\nu_n}=(n-1)\partial_{(\nu_2}\xi^{(1)\mu_1\nu_1\|}_{~~~~~~~~\,\nu_3...\nu_n)}.\eeq{gamma-1curl}
It is easy to see that, in general, an $m$-curl of Eqn.~(\ref{Gammaaction2}) gives, for $m\leq n$, \beq \Gamma\,
\psi^{(m)\mu_1\nu_1|...|\mu_m\nu_m\|}_{~~~~~~~~~~~~~~~~~~\,\nu_{m+1}...\nu_n}=(n-m)\partial_{(\nu_{m+1}}\xi^{(m)
\mu_1\nu_1|...|\mu_m\nu_m\|}_{~~~~~~~~~~~~~~~~~~\,\nu_{m+2}...\nu_{n)}}.\eeq{gamma-mcurl} In particular, when
$m=n$, we have the $\Gamma$-variation of the curvature, which vanishes: \beq \Gamma\,\Psi^{\mu_1\nu_1|...|\mu_n
\nu_n}=0.\eeq{gamma-ncurl} Notice that the $\Gamma$-closedness of the curvature does \emph{not} require any
constraints on the fermionic ghost. That the curvatures are not $\Gamma$-exact simply follows from the fact that
these are $pgh$-0 objects, whereas any $\Gamma$-exact piece must have $pgh>0$. Therefore, the curvatures are in
the cohomology of $\Gamma$, and so are their derivatives.

We have seen that only the highest curl ($n$-curl) of the spinor $\psi_{\nu_1...\nu_n}$ is $\Gamma$-closed, while
the lower curls are not. The key point is the commutativity of partial derivatives, and clearly, any arbitrary
derivative of the field will not be $\Gamma$-closed in general. However, some particular linear combination of such
objects (or $\gamma$-traces thereof) \emph{can} be $\Gamma$-closed under the \emph{constrained} ghost. The latter
possibility is exhausted by the Fronsdal tensor and its derivatives, which we will discuss later.

\subsection{The Antifields}

The antifields  $\{A^{*\mu}, C^*, \bar{\psi}^{*\mu_1...\mu_n}, \bar{\xi}^{*\mu_1...\mu_{n-1}}\}$ and their
derivatives also belong to the cohomology of $\Gamma$. Clearly, these objects are $\Gamma$-closed since $\Gamma$
does not act on the antifields, while they cannot be $\Gamma$-exact because they have $pgh=0$.

\subsection{The Ghosts \& Curls Thereof}

The \emph{undifferentiated} ghosts $\{C, \xi_{\mu_1...\mu_{n-1}}\}$ are $\Gamma$-closed objects simply because
$\Gamma$ does not act on them. Also they cannot be $\Gamma$-exact, thanks to Eqn.~(\ref{noname7}), which tells
us that any $\Gamma$-exact piece must contain at least one derivative of any of the ghosts.

Any derivatives of the ghosts will also be $\Gamma$-closed. Some derivatives, however, will be $\Gamma$-exact, and
therefore trivial in the cohomology of $\Gamma$. One can immediately dismiss as trivial \emph{any} derivative of
the bosonic ghost $C$, because $\partial_\mu C=\Gamma A_\mu$ from Eqn.~(\ref{Gammaaction1}).

Derivatives of the fermionic ghost\,\footnote{The rest of this Appendix will deal only with the fermionic ghost
$\xi_{\mu_1...\mu_n}$, and without any source of confusion we will simply call it ``ghost''.} are more subtle.
One can show that only the $\gamma$-traceless part of the curls of the ghost $\{\xi^{(m)}_{\mu_1\nu_1|...|
\mu_m\nu_m\|\nu_{m+1}...\nu_{n-1}},\, m\leq n-1\}$ are non-trivial elements in the cohomology of $\Gamma$.
First, one can convince oneself step by step why only the ghost-curls are interesting. In the simplest
non-trivial case of $n=2$, we see that \beq \partial_\mu\xi_\nu=\partial_{(\mu}\xi_{\nu)}+\partial_{[\mu}
\xi_{\nu]}=\tfrac{1}{2}\Gamma\psi_{\mu\nu}+\tfrac{1}{2}\xi^{(1)}_{\mu\nu}.\eeq{gamma2} For $n=3$, we find
\beq \partial_\mu\xi_{\nu\rho}=\partial_{(\mu}\xi_{\nu\rho)}+\tfrac{4}{3}\partial_{[\mu} \xi_{\nu]\rho}
+\tfrac{2}{3}\partial_{[\nu}\xi_{\rho]\mu}=\tfrac{1}{3}\Gamma \psi_{\mu\nu\rho}+\tfrac{2}{3}\xi^{(1)}_{\mu
\nu\|\rho}+\tfrac{1}{3}\xi^{(1)}_{\nu\rho\|\mu},\eeq{gamma3} It is easy to generalize this to the arbitrary
spin case, for which we obtain \bea \partial_\rho\xi_{\nu_1...\nu_{n-1}}&=&\partial_{(\rho}\xi_{\nu_1...
\nu_{n-1})}+2\left(1-\tfrac{1}{n}\right)\partial_{[\rho}\xi_{\nu_1]\nu_2...\nu_{n-1}}\nonumber\\&&+2\sum
_{m=1}^{n-2}\left(1-\tfrac{m+1}{n}\right)\partial_{[\nu_m}\xi_{\nu_{m+1}]\rho\,\nu_1...\nu_{m-1}\nu_{m+2}...
\nu_{n-1}}\nonumber\\&=&\tfrac{1}{n}\,\Gamma\psi_{\rho\,\nu_1...\mu_{n-1}}+\left(1-\tfrac{1}{n}\right)\xi_{\rho
\nu_1\|\nu_2...\nu_{n-1}}^{(1)}\nonumber\\&&+\sum_{m=1}^{n-2}\left(1-\tfrac{m+1}{n}\right)\xi_{\nu_m\nu_{m+1}
\|\rho\,\nu_1...\nu_{m-1}\nu_{m+2}...\nu_{n-1}}^{(1)}.\eea{gamma4} In view of Eqn.~(\ref{gamma2})--(\ref{gamma4}),
we conclude that any first derivative of the ghost is a linear combination of 1-curls, up to $\Gamma$-exact
terms. Therefore, in the cohomology of $\Gamma$ it suffices to consider only 1-curls of the ghost. More
generally, one can consider only the $m$-curls in the cohomology of $\Gamma$, instead of arbitrary $m$
derivatives, where $m\leq n-1$. The latter can easily be seen by first taking a 1-curl of Eqn.~(\ref{gamma4}),
and convincing oneself that only 2-curls of the ghost are interesting. In this way, one can continue up to
$m$-curls of Eqn.~(\ref{gamma4}), $m\leq n-1$, to show that it suffices to consider only $m$-curls.

The derivative of an $m$-curl, $\partial_{\nu_n}\xi^{(m)}_{\mu_1\nu_1|...|\mu_m\nu_m\|\,\nu_{m+1}...\nu_{n-1}}$,
contains non-trivial $(m+1)$-curls plus trivial terms. It is clear that this quantity can be $\Gamma$-exact
if and only if symmetrized w.r.t. the indices $\{\nu_{m+1}, ..., \nu_n\}$. In this case, we have from
Eqn.~(\ref{gamma-mcurl}) \beq \partial_{(\nu_n}\xi^{(m)\mu_1\nu_1|...|\mu_m\nu_m\|}_{~~~~~~~~~~~~~~~~~~\,
\nu_{m+1}...\nu_{n-1})}=\tfrac{1}{n-m}\,\Gamma\,\psi^{(m)\mu_1\nu_1|...|\mu_m\nu_m\|}_{~~~~~~~~~~~~~~~~~~\,
\nu_{m+1}...\nu_n},\qquad 0\leq m\leq n-1.\eeq{exc} It follows immediately that a derivative of the highest
ghost-curl is always $\Gamma$-exact. Indeed, in Eqn.~(\ref{exc}) one can set $m=n-1$, and obtain
\beq \partial_{\nu_n}\xi^{(n-1)}_{\mu_1\nu_1|...|\mu_{n-1}\nu_{n-1}}=\Gamma\,\psi^{(n-1)}_{\mu_1\nu_1|...|
\mu_{n-1}\nu_{n-1}\|\,\nu_n}.\eeq{exc1}

Although any $m$-curl, $\xi^{(m)}_{\mu_1\nu_1|...|\mu_m\nu_m||\nu_{m+1}...\nu_{n-1}}$, is in the cohomology
of $\Gamma$, its $\gamma$-trace is always $\Gamma$-exact. In fact, the latter vanishes when the $\gamma$-matrix
to be contracted carries one of the unpaired indices $\{\nu_{m+1},...,\nu_{n-1}\}$, thanks to the
$\gamma$-tracelessness of the ghost. If the index contraction is otherwise, the same constraint gives
\beq \gamma^{\mu_1}\xi^{(m)}_{\mu_1\nu_1|...|\mu_m\nu_m\|\nu_{m+1}...\nu_{n-1}}=\displaystyle{\not{\!\partial}}
\,\xi^{(m-1)}_{\mu_2\nu_2|...|\mu_m\nu_m\|\nu_1\nu_{m+1}...\nu_{n-1}}.\eeq{gamma5} On the other hand, one can
take a $\gamma$-trace of Eqn.~(\ref{exc}) to have \beq \Gamma\,\displaystyle{\not{\!\psi}}^{(m)}_{\mu_1\nu_1|
...|\mu_m\nu_m\|\,\nu_{m+1}...\nu_{n-1}}=(n-m)\displaystyle{\not{\!\partial}}\,\xi^{(m-1)}_{\mu_2\nu_2|...|
\mu_m\nu_m\|\nu_1\nu_{m+1}...\nu_{n-1}}.\eeq{gamma6} Comparing Eqn.~(\ref{gamma5}) with~(\ref{gamma6}),
it is clear that $\gamma^{\mu_1}\xi^{(m)}_{\mu_1\nu_1|...|\mu_m\nu_m\|\nu_{m+1}...\nu_{n-1}}$ is $\Gamma$-exact.
Therefore, one can exclude the $\gamma$-traces of ghost curls from the cohomology of $\Gamma$.

\subsection{The Fronsdal Tensor}

The Fronsdal tensor $\mathcal S_{\mu_1...\mu_n}$ and its derivatives are also in the cohomology of $\Gamma$.
From the definition~(\ref{Fronsdal}), $\mathcal S_{\mu_1...\mu_n}$ can be shown to be $\Gamma$-closed under
the constrained ghost. Indeed, \bea \Gamma\mathcal{S}_{\mu_1...\mu_n}&=&i\left[\displaystyle{\not{\!\partial
\,}}\Gamma \psi_{\mu_1...\mu_n}-n\partial_{(\mu_1}\Gamma\displaystyle{\not{\!\psi\,}}_{\mu_2...\mu_n)}\right]
\nonumber\\&=& in\left[\displaystyle{\not{\!\partial}}\,\partial_{(\mu_1}\xi_{\mu_2...\mu_n)}-n\gamma^\rho
\partial_{(\mu_1}\partial_{(\rho}\xi_{\mu_2...\mu_n))}\right]\nonumber\\&=&-in(n-1)\partial_{(\mu_1}\partial
_{(\mu_2}\displaystyle{\not{\!\xi}}_{\mu_3\dots\mu_n))},\eea{gamma7} which vanishes if the ghost is
$\gamma$-traceless. Being a $pgh$-0 object, $\mathcal S_{\mu_1...\mu_n}$  can also not be $\Gamma$-exact.
So, the Fronsdal tensor and its derivatives belong to the cohomology of $\Gamma$.

However, in view of Eqn.~(\ref{eomarb2}) and~(\ref{eomarb4}), the two highest curls of the Fronsdal tensor
boil down to objects we have already enlisted in Subsection~\ref{sec:Curvatures}, and therefore need not be
considered separately. These equations are generalizations of the Damour-Deser relations~\cite{DD} (see 
also~\cite{BB}). For spin $\tfrac{5}{2}$, in particular, they make it sufficient to consider only symmetrized 
derivatives of the Fronsdal tensor.

\end{appendix}

\end{document}